\documentclass[a4paper,10pt]{article}
\usepackage[T2A]{fontenc}
\usepackage[utf8]{inputenc}
\usepackage[english, russian]{babel}
\usepackage{amssymb,amsmath}
\usepackage{booktabs}
\usepackage{multirow}
\usepackage{graphicx}
\usepackage{color}
\usepackage[mathlines]{lineno}% Enable numbering
\linenumbers\par %%% <---- turn on the numeration of lines
\usepackage{comment}
\graphicspath{{images/}}
\usepackage{indentfirst} \righthyphenmin=2 %????????? ??????? ???? ????
\usepackage[pdftex,unicode,colorlinks, citecolor=blue,%
filecolor=black, linkcolor=blue, urlcolor=black]{hyperref}
\usepackage[figure,table]{hypcap} %?????? ?? ?????? ???????, ???????...
\begin{document}
\def\slantfrac#1#2{ \hspace{3pt}\!^{#1}\!\!\hspace{1pt}/
  \hspace{2pt}\!\!_{#2}\!\hspace{3pt} }

\author{Dmitry~L.~Markovich$^{1*}$, Pavel~Ginzburg$^{2}$$^{**}$,
  Anton~Samusev$^{1}$, \\Pavel~A.~Belov$^{1}$, and
  Anatoly~V.~Zayats$^2$}

\title{Magnetic dipole radiation tailored by substrates: numerical
  investigation}

\date{}
\maketitle

\renewcommand{\abstractname}{}
\begin{center}
  {$^1$ St.~Petersburg National Research University of Information
    Technologies, Mechanics and Optics, 49 Kronverskii Ave.,
    St.~Petersburg 197101, Russian Federation}

  {$^2$ Department of Physics, King's College London, Strand, London
    WC2R 2LS, UK}
\end{center}

\begin{center}
  Corresponding author: *dmmrkovich@gmail.com, **pavel.ginzburg@kcl.ac.uk
\end{center}

\begin{abstract}
  Nanoparticles of high refractive index materials can possess strong
  magnetic polarizabilities and give rise to artificial magnetism in
  the optical spectral range. While the response of individual
  dielectric or metal spherical particles can be described
  analytically via multipole decomposition in the Mie series, the
  influence of substrates, in many cases present in experimental
  observations, requires different approaches. Here, the comprehensive
  numerical studies of the influence of a substrate on the spectral
  response of high-index dielectric nanoparticles were performed. In
  particular, glass, perfect electric conductor, gold, and hyperbolic
  metamaterial substrates were investigated. Optical properties of
  nanoparticles were characterized via scattering cross-section
  spectra, electric field profiles, and induced electric and magnetic
  moments. The presence of substrates was shown to introduce
  significant impact on particle’s magnetic resonances and resonant
  scattering cross-sections. Variation of substrate material provides
  an additional degree of freedom in tailoring properties of emission
  of magnetic multipoles, important in many applications.
\end{abstract}
%\renewcommand{\contentsname}{Contents}
% \tableofcontents{}
\section{Introduction}
\label{introduction}
Materials can respond to applied electromagnetic field via both
electric and magnetic susceptibilities. While dielectrics demonstrate
relatively high values of permittivities even at high frequencies
(ultraviolet spectral range or lower), inherent natural permeabilities
rapidly approach unity, having certain high-frequency cut-off around
MHz frequencies~\cite{polydoroff-magnetic-1960}. This natural property
results from relatively fast electronic polarizabilities and
electronic band structure that has resonances at optical frequencies
and slow spin and orbital interactions which define the magnetic
susceptibility. Nevertheless, optical magnetism can be artificially
created via carefully engineered subwavelength structures. For
example, arrays of ordered split-ring resonators, made of conducting
metals, can create artificial magnetic responses at high frequencies
(THz range) and even produce negative effective
permeabilities~\cite{smith-composite-2000}. Similar concepts can be
also employed in optics. With proper choice of shape of metallic
nanostructures and their arrangements, called metamaterials,
artificial magnetism has been demonstrated in the infra-red and
visible spectral range~\cite{zheludev-the-road-2010,
  soukoulis-optical-2010, boltasseva-low-loss-2011,
  shalaev-optical-2007}.  One of the fundamental bottlenecks, limiting
the performance of plasmonic components and metamaterials is inherent
material losses that is of great importance for various nanophotonic
components, such as nanolenses~\cite{radko-surface-2007,
  ginzburg-plasmonic-2011}, antennas~\cite{curto-unidirectional-2010},
particle-based waveguides~\cite{zayats-nano-optics-2005}, ordered
particle arrays~\cite{maier-observation-2002}, and
biosensors~\cite{anker-biosensing-2008}.  At the same time, high
refractive index dielectric nanoparticles have shown to be promising
in the context of artificial magnetism and the majority of
aforementioned components can be implemented with all-dielectric
elements, as was already demonstrated in case of ordered particle
arrays~\cite{evlyukhin-optical-2010} and antenna
applications~\cite{krasnok-all-dielectric-2012}. Resonant phenomena in
positive permittivity particles, in contrary to subwavelength
plasmonic structures, rely on the retardation effects. Nevertheless,
high index spherical particles of nanometric dimensions can exhibit
multiple resonances in the visible spectral range. In particular,
strong resonant light scattering associated with the excitation of
magnetic and electric dipolar modes in silicon nanoparticles has
recently been demonstrated experimentally using dark-field optical
microscopy~\cite{evlyukhin-demonstration-2012,
  kuznetsov-magnetic-2012} and directional light scattering by
spherical silicon nanoparticles in the visible spectral range has been
reported~\cite{fu-directional-2013}. While the majority of theoretical
studies consider isolated spherical particles or their clusters in
free space, very often experimental geometries involve the presence of
substrates where nanoparticles are
placed~\cite{evlyukhin-demonstration-2012, kuznetsov-magnetic-2012,
  fu-directional-2013}. Therefore, investigations of substrate effects
on magnetic dipole resonances in dielectric nanoparticles are of great
importance for understanding both fundamental phenomena and
predictions of experimental measurements. In this paper, we performed
numerical studies of optical properties of high-index dielectric
nanoparticles on various types of substrates. In particular, glass,
gold and hyperbolic metamaterial substrates were considered. Optical
properties of nanoparticles were characterized via scattering
cross-section spectra, electric field profiles, and induced electric
and magnetic moments.

\section{Theoretical and numerical frameworks}
\renewcommand{\figurename}{Fig.}
\begin{figure}[h!]
  \begin{minipage}[h]{0.99\textwidth}
    \begin{center}
      \includegraphics[width=0.2\textwidth]{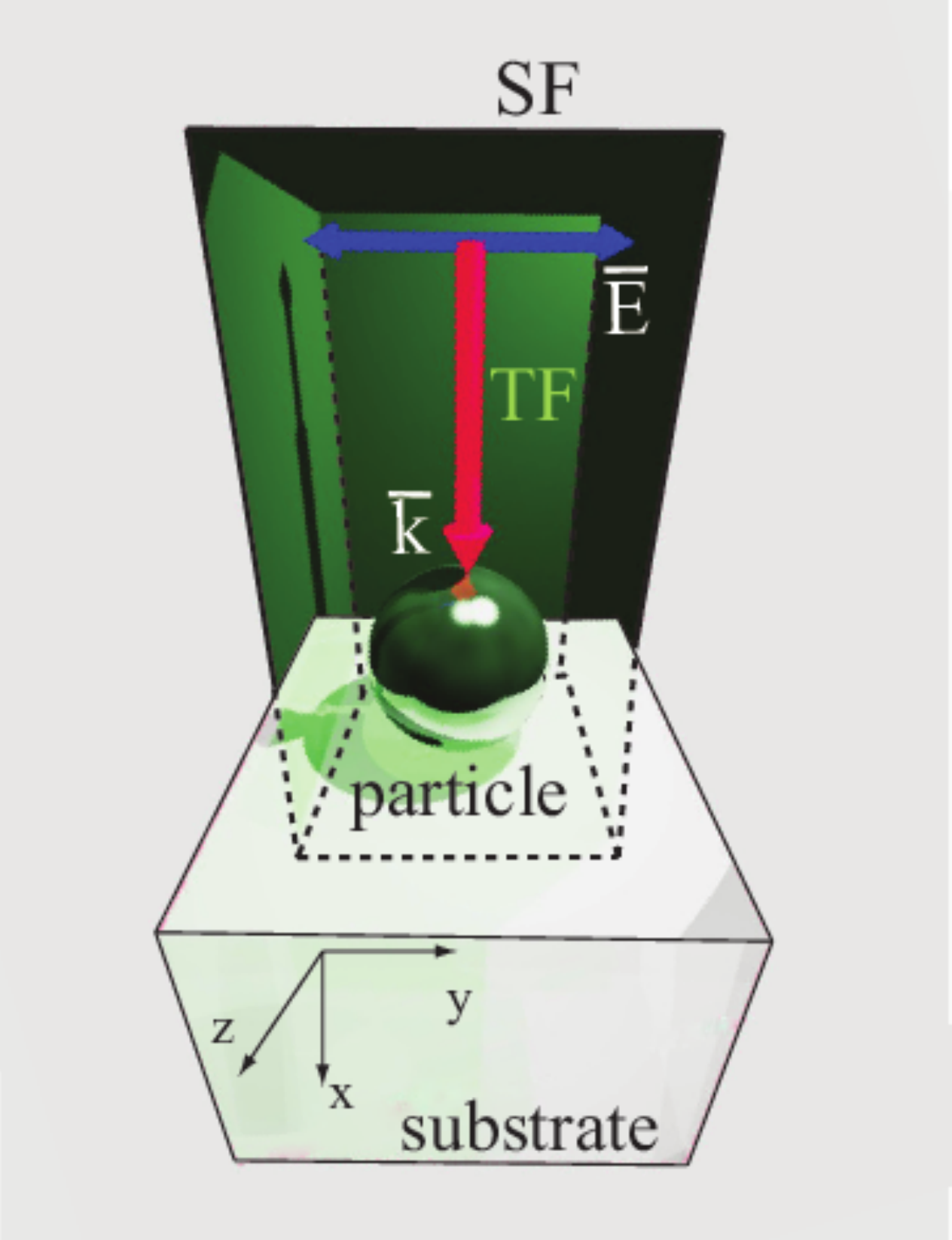}
    \end{center}
  \end{minipage}
  \caption{Dielectric nanoparticle on a substrate illuminated with a
    plane wave. The regions inside and outside the green box depicts
    total (incident and scattered) electric field (TF) and scattered
    electric field (SF), respectively.}
  \label{geometry_01}
\end{figure}

Optical response of individual spherical particles embedded in
homogeneous host materials can be described analytically via the Mie
series decomposition~\cite{bohren-absorption-1998}. Generally, any
shape with boundaries belonging to ``coordinate surfaces'' sets at
coordinate systems where Helmholtz equation is separable, i.e., has
analytic solution for scattering problems. However, the separation of
variables method breaks down once substrates are introduced. As the
first order approximation, the image theory could be
employed~\cite{hammond-electric-1960, fan-near-normal-2012,
  berkovitch-nano-plasmonic-2012}, while more complex treatments can
account for retardation effects and higher multi-pole
interactions~\cite{moreno-light-2007, pena-sizing-1999,
  bobbert-light-1986}. Considerations of anisotropic substrates lead
to major complexity even in the image theory (point charge images
should be replaced by certain
distributions)~\cite{lindell-electrostatic-2004} and make fully
analytical approaches to be of limited applicability. Moreover, in the
case of high-index dielectric particles, the resonances are
overlapping in frequency, making even the image theory to be
inapplicable. At the same time, this spectral overlap can be employed
for super-directive antenna
applications~\cite{kuznetsov-magnetic-2012,
  krasnok-superdirective-2013}.

One of the commonly used techniques for numerical analysis of
scattering processes is the so-called "total-field scattered-field"
(TFSF) approach~\cite{taflove}. The key advantage of this method is
the separation of relatively weak scattered field (SF) from
predominating high amplitude total field (TF) which contains both
incident and scattered fields, in a distinct simulation domain. The
TFSF method also allows to subtract the electric field, reflected
backwards by the substrate in the SF domain, enabling the calculation
of a scattering cross-section.

The geometrical arrangement of the considered scenario is represented
in Fig.~\ref{geometry_01}: a small dielectric spherical particle (r =
70 nm, $\epsilon$ = 20) is placed on the substrate. The centre of the
particle coincides with the coordinate origin.  The system is
illuminated with a short pulse (in time) with broadband spectrum
covering the spectral range from 400 to 750 nm. FDTD method allows
analyzing spectral responses via single simulation by adopting the
Fourier decomposition method with subsequent normalization. The key
parameters characterizing the system are the electric and magnetic
dipole moments defined as
\begin{eqnarray}
  {\bf p} = \epsilon_0 \iiint\limits_{V_{sphere}} \! (\epsilon(r, \omega) - 1)
  {\bf E}(r, \omega)~\mathrm{d}V \nonumber \\
  {\bf m} = \frac{i \omega \cdot \epsilon_0}{2} \iiint\limits_{V_{sphere}} \! (\epsilon(r, \omega) - 1)
  {\bf E}(r, \omega) \times {\bf r}~\mathrm{d}V,
  \label{pm_calculation}
\end{eqnarray}
where $\epsilon_0$ is the vacuum permittivity, $V_{sphere} \simeq$
1.44e$^{-3}$ $\mu$m$^3$ is the volume of the particle,
$\epsilon(r,\omega)$ is the frequency-dependent particle permittivity,
$\omega$, is the frequency of the incident light, and ${\bf E}(r,
\omega)$ is the electric field in frequency domain.
Generally, these dipolar moments in the asymmetric system considered
here depend on the illumination polarization, direction of incidence
and spatial shape of the beam. In the following, a normally incident
y-polarized plane wave was considered. To avoid the interplay between
geometrical parameters and chromatic dispersion of the particle’s
material, the latter was neglected.
\clearpage
\section{Results and discussion}
\subsection{Optical properties of a dielectric nanoparticle in free space}
\begin{figure}[h!]
  \begin{minipage}[h]{0.49\textwidth} a) \end{minipage} \hfill
  \begin{minipage}[h]{0.49\textwidth} b) \end{minipage} \vfill
  \begin{minipage}[h]{0.49\textwidth}
    \begin{center}
      \includegraphics[width=0.99\textwidth]{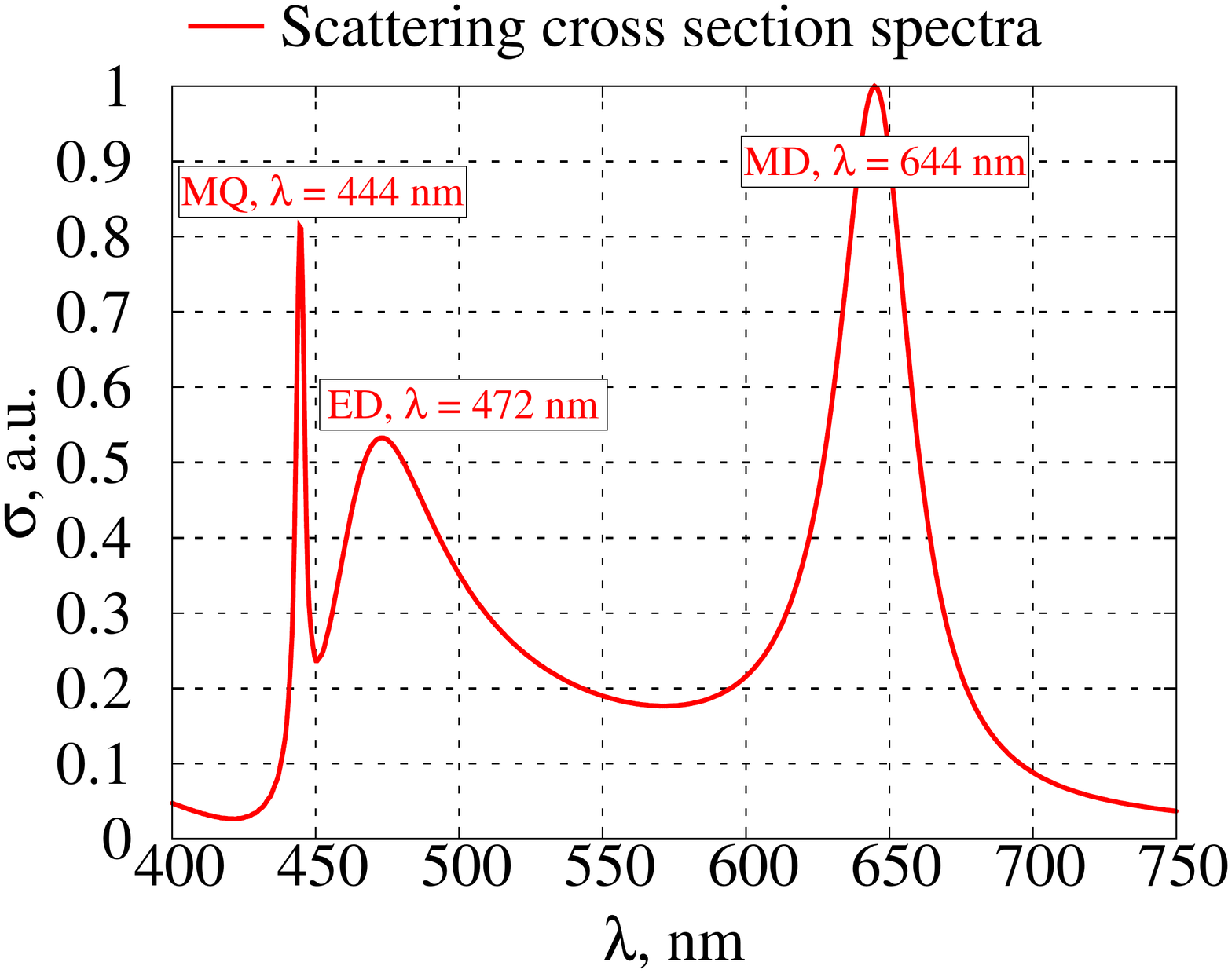}
    \end{center}
  \end{minipage}
  \hfill
  \begin{minipage}[h]{0.49\textwidth}
    \begin{center}
      {\normalfont
        \begin{tabular}{cccc}
          \toprule
          \multicolumn{4}{c}{Particle in free space} \\
          \midrule
          $\lambda$, [nm] & 444 & 472 & 644\\
          \midrule
          $\sigma$ / S & 2.8 & 1.8 & 3.4\\
          \midrule
          $p_y$, [e $\cdot$ nm]& 1.6$e^{-4}$ & 1.0$e^{-3}$ & 7.7$e^{-4}$\\
          \midrule
          $m_z$, [A $\cdot$ nm$^2$]& 1.5$e^{-3}$ &1.8$e^{-3}$ & 2.1$e^{-2}$\\
          \bottomrule
        \end{tabular}
      }
    \end{center}
  \end{minipage}
  \vfill
  \begin{minipage}[h]{0.3\textwidth} c) \end{minipage} \hfill
  \begin{minipage}[h]{0.3\textwidth} d) \end{minipage} \hfill
  \begin{minipage}[h]{0.3\textwidth} e) \end{minipage} \vfill
  \begin{minipage}[h]{0.3\textwidth}
    \begin{center}
      \includegraphics[width=0.99\textwidth]
                      {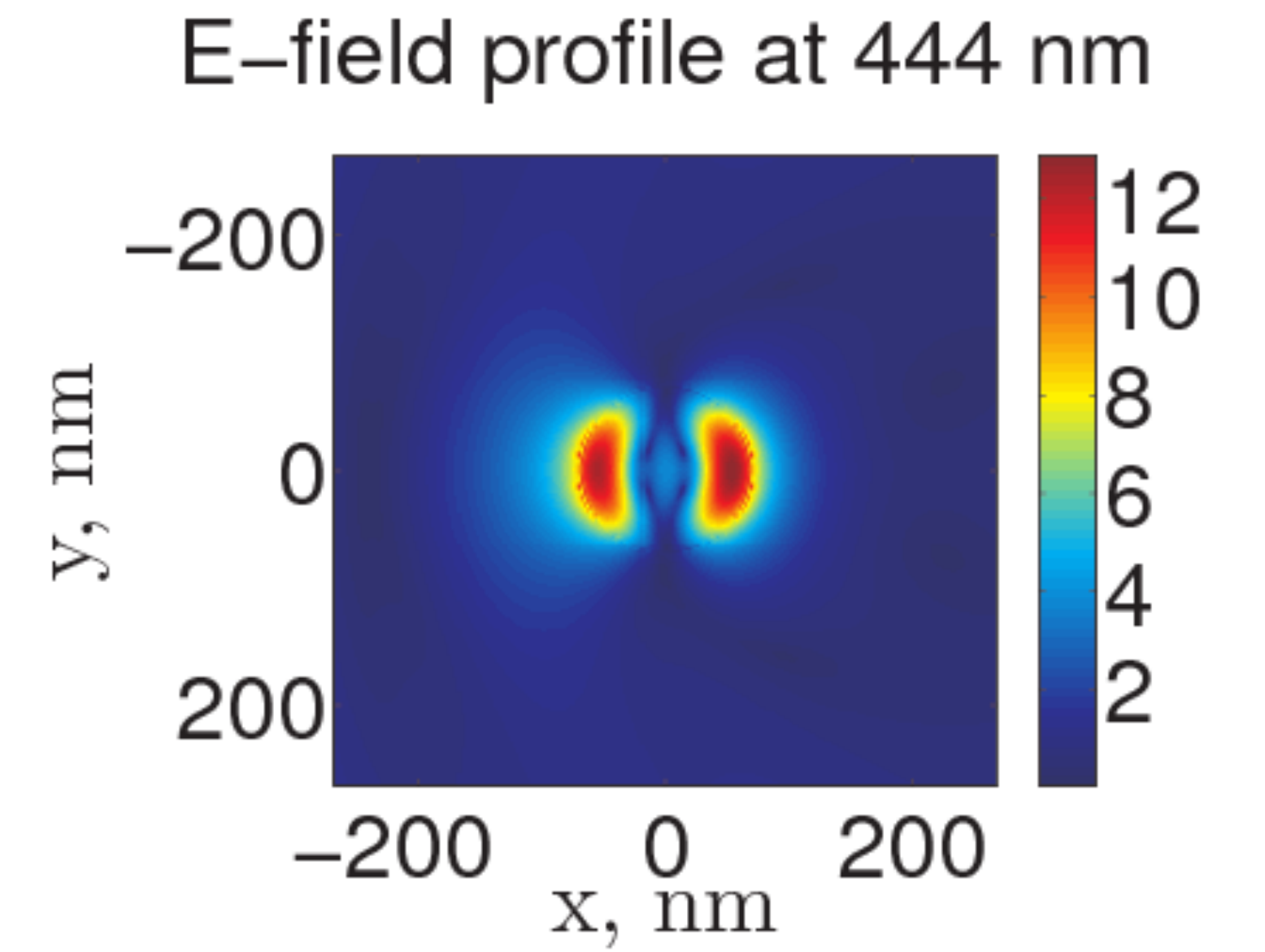}
    \end{center}
  \end{minipage}
  \hfill
  \begin{minipage}[h]{0.3\textwidth}
    \begin{center}
      \includegraphics[width=0.99\textwidth]
                      {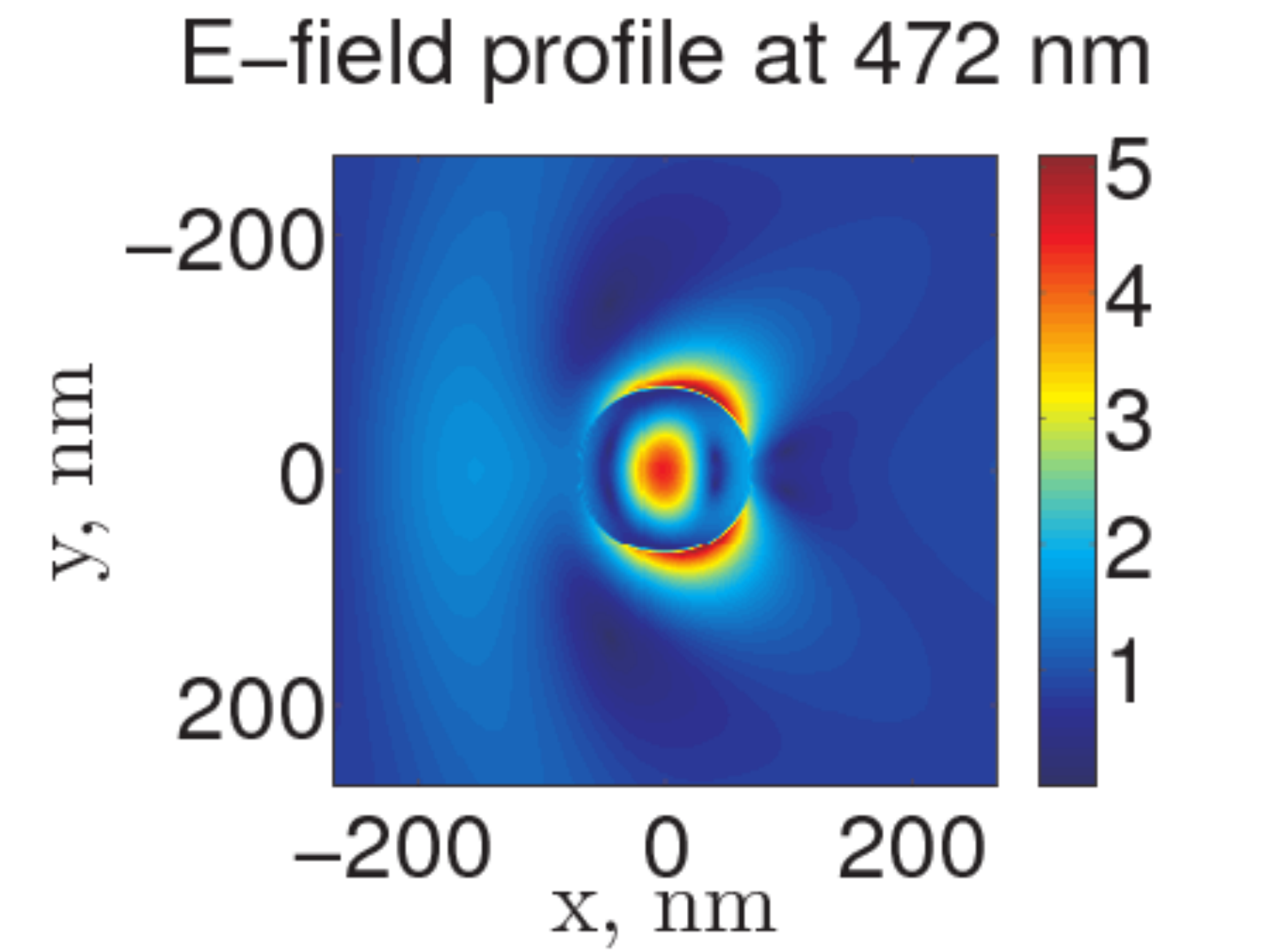}
    \end{center}
  \end{minipage}
  \hfill
  \begin{minipage}[h]{0.3\textwidth}
    \begin{center}
      \includegraphics[width=0.99\textwidth]
                      {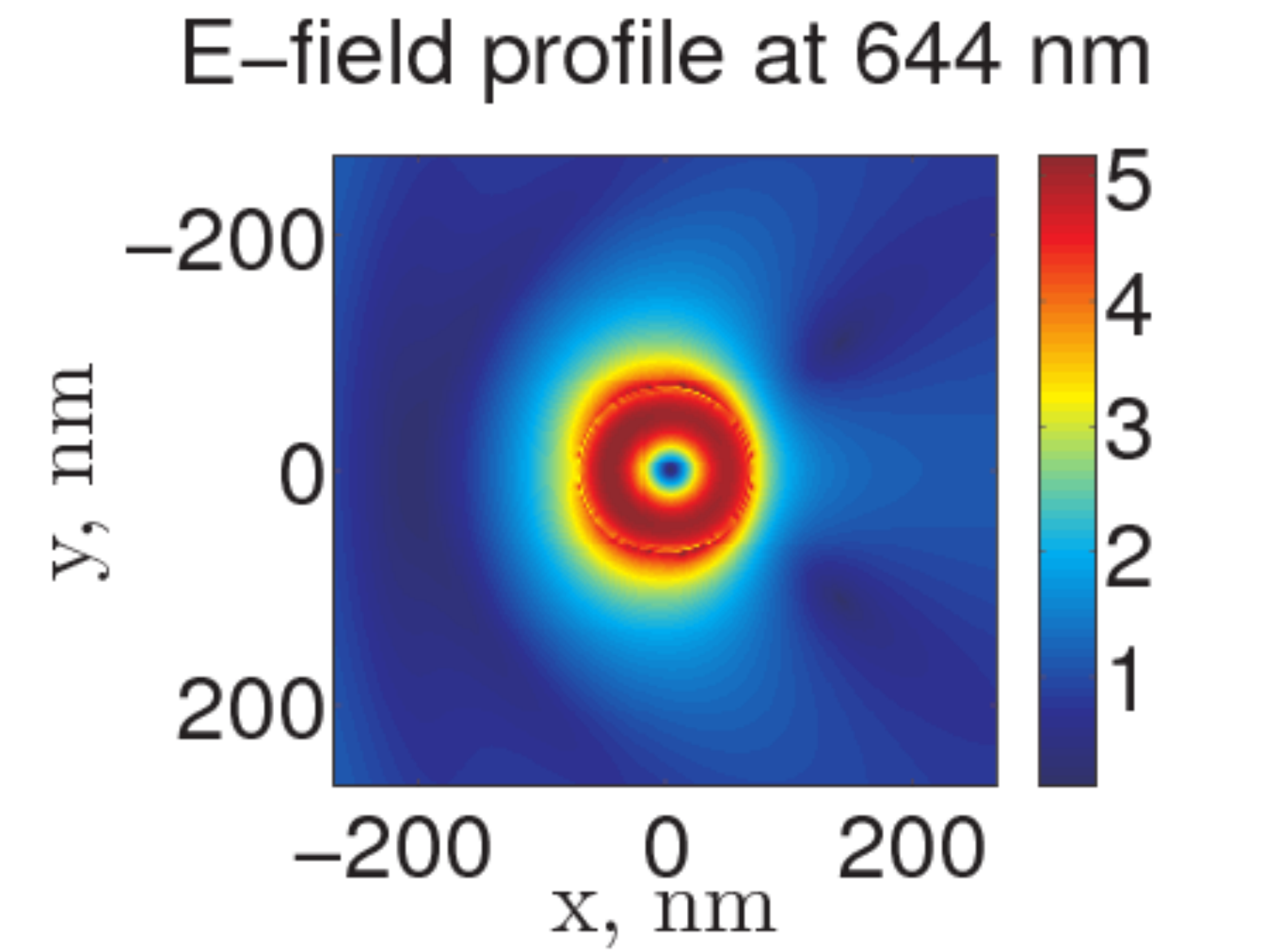}
    \end{center}
  \end{minipage}
  \vfill
\caption{Optical properties of a dielectric ($\epsilon_{particle} =
  20$) nanoparticle of 70 nm radius in free space: (a) scattering
  cross-section spectra, (b) the summary of the resonant wavelengths,
  the scattering cross-section normalised to the geometric
  cross-section, and electric $p_y$ and magnetic moments $m_z$, (c-e)
  the spatial distribution of the electric field amplitudes at
  resonant wavelengths (normalised to the incident
  field).}
\label{vacuum_substrate}
\end{figure}

To verify the accuracy of the approach, optical properties of the
particle in free space were evaluated numerically and compared to
analytic description. Scattering cross-section spectra
(Fig.~\ref{vacuum_substrate}(a)) shows 3 distinctive resonances
corresponding to magnetic dipole (MD), electric dipole (ED), and
magnetic quadrupole (MQ) at 644 nm, 472 nm, and 444 nm, respectively.
The corresponding electric field amplitude distributions are depicted
on Fig.~\ref{vacuum_substrate}(c, d, and e). They show the
amplification of near field amplitude with respect to the input source
field and provide visual hints for identification of the type of a
resonance. The circular-like electric field distribution is observed
for a MD resonance (Fig.~\ref{vacuum_substrate}e), the drop-shaped
distribution for an ED resonance (Fig.~\ref{vacuum_substrate}d), and
the double resonant sectors for a MQ resonance
(Fig.~\ref{vacuum_substrate}c).

Knowing the electric field distribution in the entire space allows to
calculate the moments defined in Eq.~\ref{pm_calculation}. These
results are summarized in Fig.~\ref{vacuum_substrate}(b). Since the
particle is illuminated along the x-axis with the y-polarized plane
wave, the electric moment component $p_y$ and magnetic moment
component $m_z$ are expected to be dominating. Having calculated all
other components for both electric and magnetic moments, we found
$p_y$ and $m_z$ to be more than 1$e^{17}$ times larger than other
components, confirming symmetry arguments. It is instructive to
compare the obtained values to the induced electric dipolar moment of
a gold sphere of the same radius~\cite{maier-plasmonics-2007} which is
$p_{gold}$ = 2.58$e^{-4}$ [e$\cdot$nm] at 472 nm; hence, a high-index
dielectric particle has $\approx$ 4 times larger electric dipolar
moment and no associated material losses. Numerically calculated
dipolar moments were compared with analytical Mie theory approach
reported in~\cite{evlyukhin-optical-2010} and differ by no more than
5\%, verifying the accuracy of the employed numerical method.

\clearpage
\subsection{Dielectric substrate}
\begin{figure}[h!]
  \begin{minipage}[h]{0.49\textwidth} a) \end{minipage} \hfill
  \begin{minipage}[h]{0.49\textwidth} b) \end{minipage} \vfill
  \begin{minipage}[h]{0.49\textwidth}
    \begin{center}
      \includegraphics[width=0.99\textwidth]{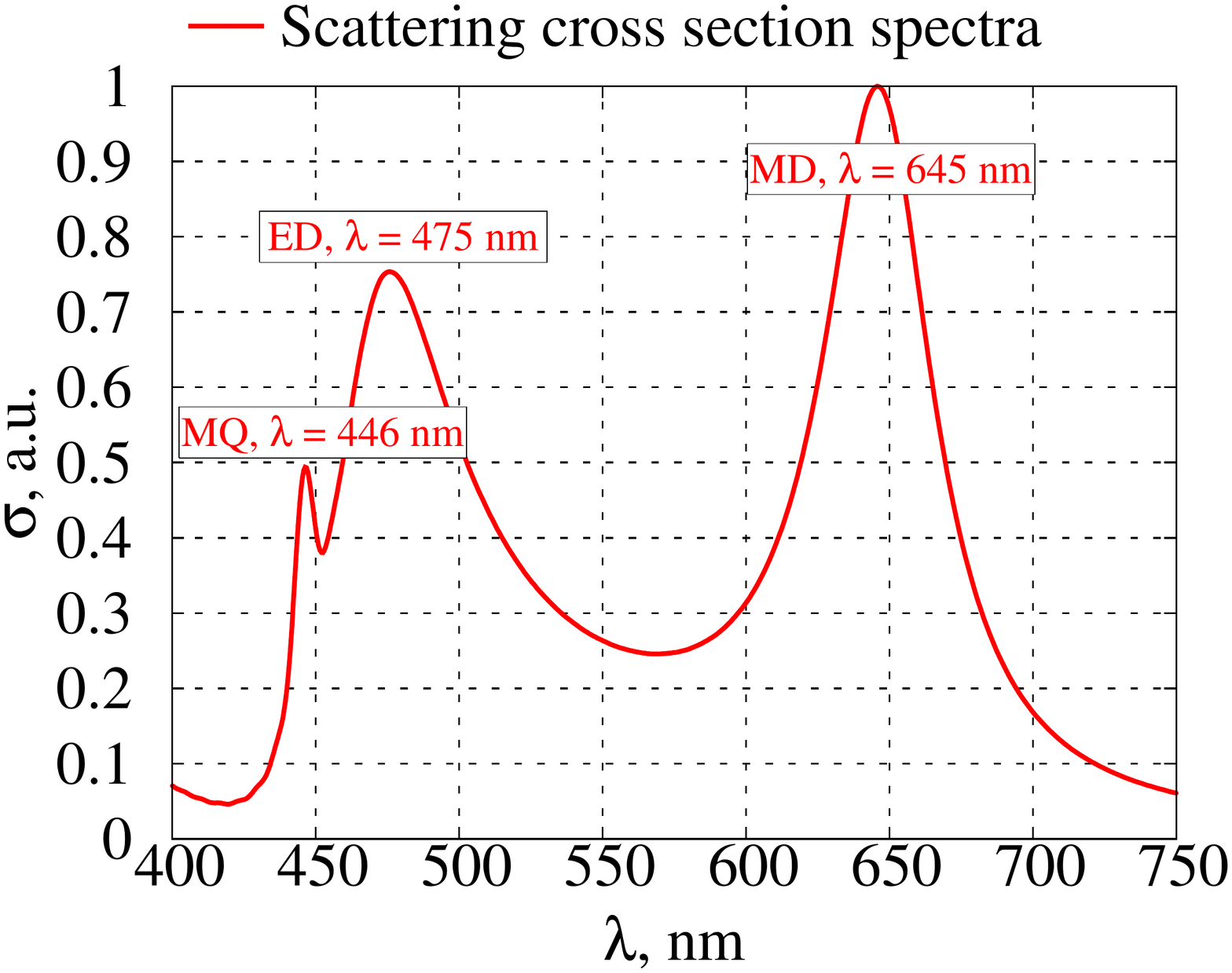}
    \end{center}
  \end{minipage}
  \hfill
  \begin{minipage}[h]{0.49\textwidth}
    \begin{center}
      {\normalfont
        \begin{tabular}{cccc}
          \toprule
          \multicolumn{4}{c}{Particle on a dielectric substrate} \\
          \midrule
          $\lambda$, nm & 446 & 475 & 645\\
          \midrule
          $\sigma$ / S & 1.5 & 2.3 & 3.0\\
          \midrule
          $p_y$, [e $\cdot$ nm] & 2.44e$^{-4}$ & 1.1$e^{-3}$ & 9.8$e^{-4}$\\
          \midrule
          $m_z$, [A $\cdot$ nm$^2$] & 1.3$e^{-3}$ & 1.2$e^{-3}$ & 1.5$e^{-2}$\\
          \bottomrule
        \end{tabular}
        }
    \end{center}
  \end{minipage}
  \vfill
  \begin{minipage}[h]{0.3\textwidth} c) \end{minipage} \hfill
  \begin{minipage}[h]{0.3\textwidth} d) \end{minipage} \hfill
  \begin{minipage}[h]{0.3\textwidth} e) \end{minipage} \vfill
  \begin{minipage}[h]{0.3\textwidth}
    \begin{center}
      \includegraphics[width=0.99\textwidth]
                      {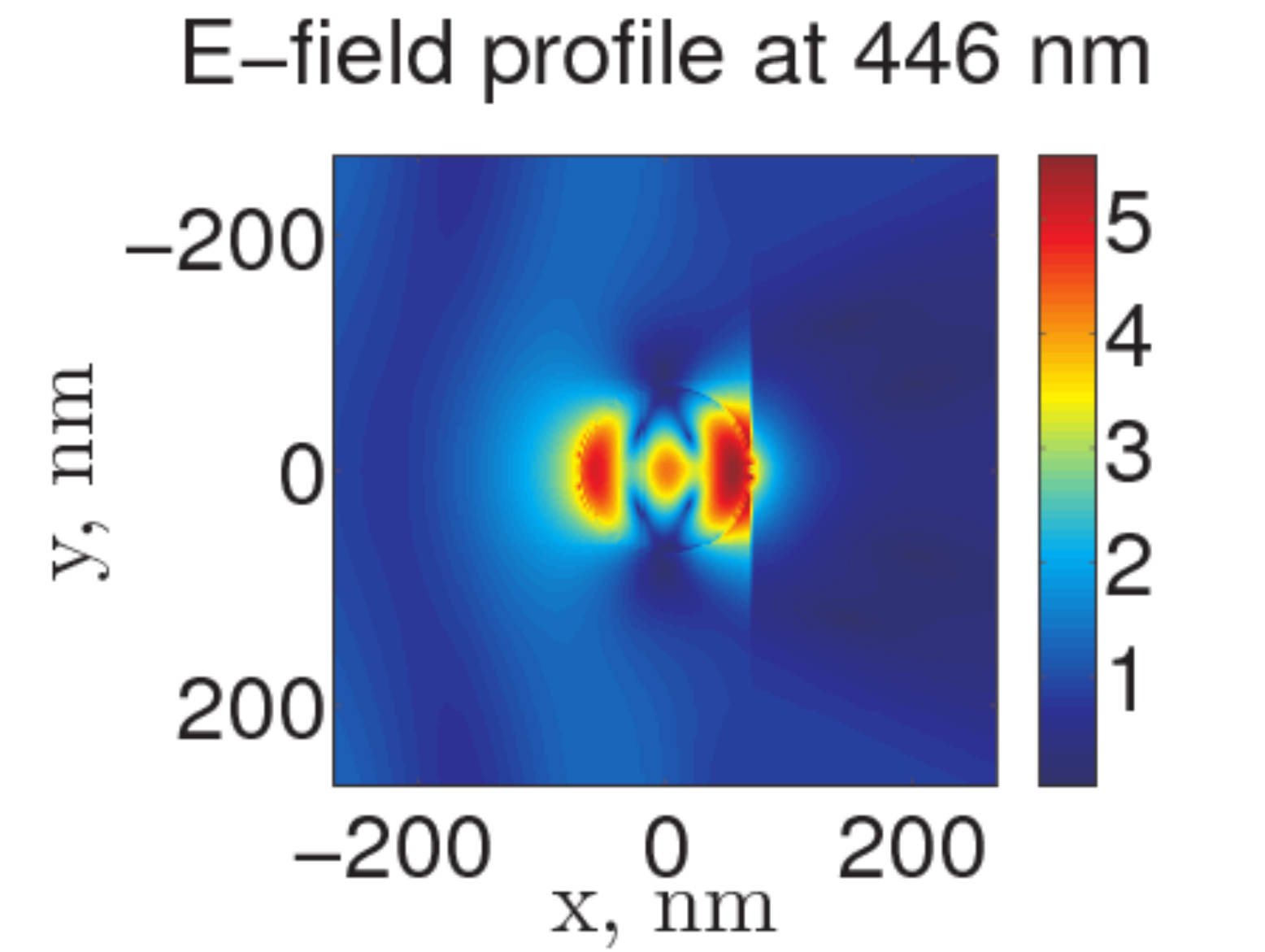}
    \end{center}
  \end{minipage}
  \hfill
  \begin{minipage}[h]{0.3\textwidth}
    \begin{center}
      \includegraphics[width=0.99\textwidth]
                      {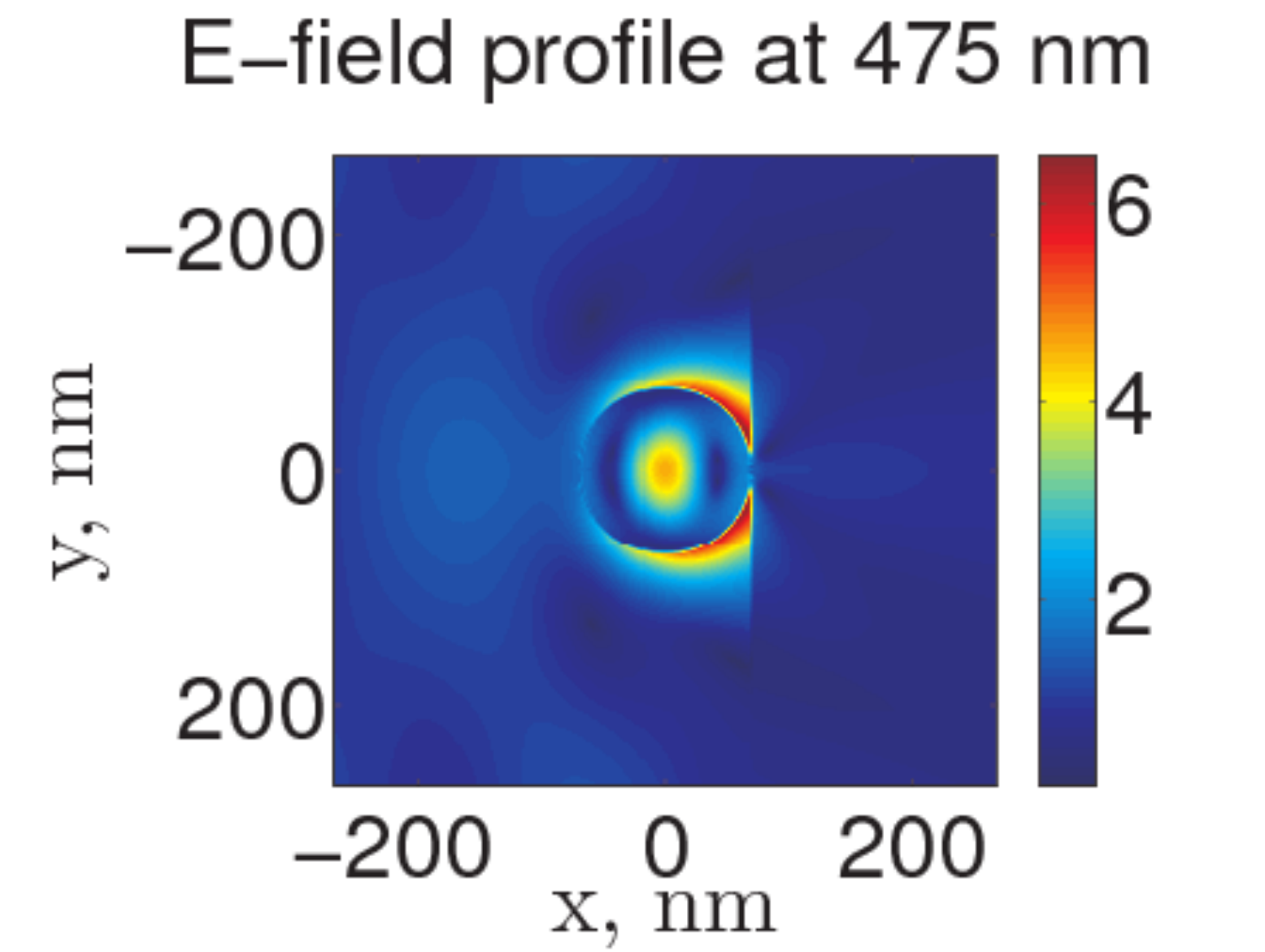}
    \end{center}
  \end{minipage}
  \hfill
  \begin{minipage}[h]{0.3\textwidth}
    \begin{center}
      \includegraphics[width=0.99\textwidth]
                      {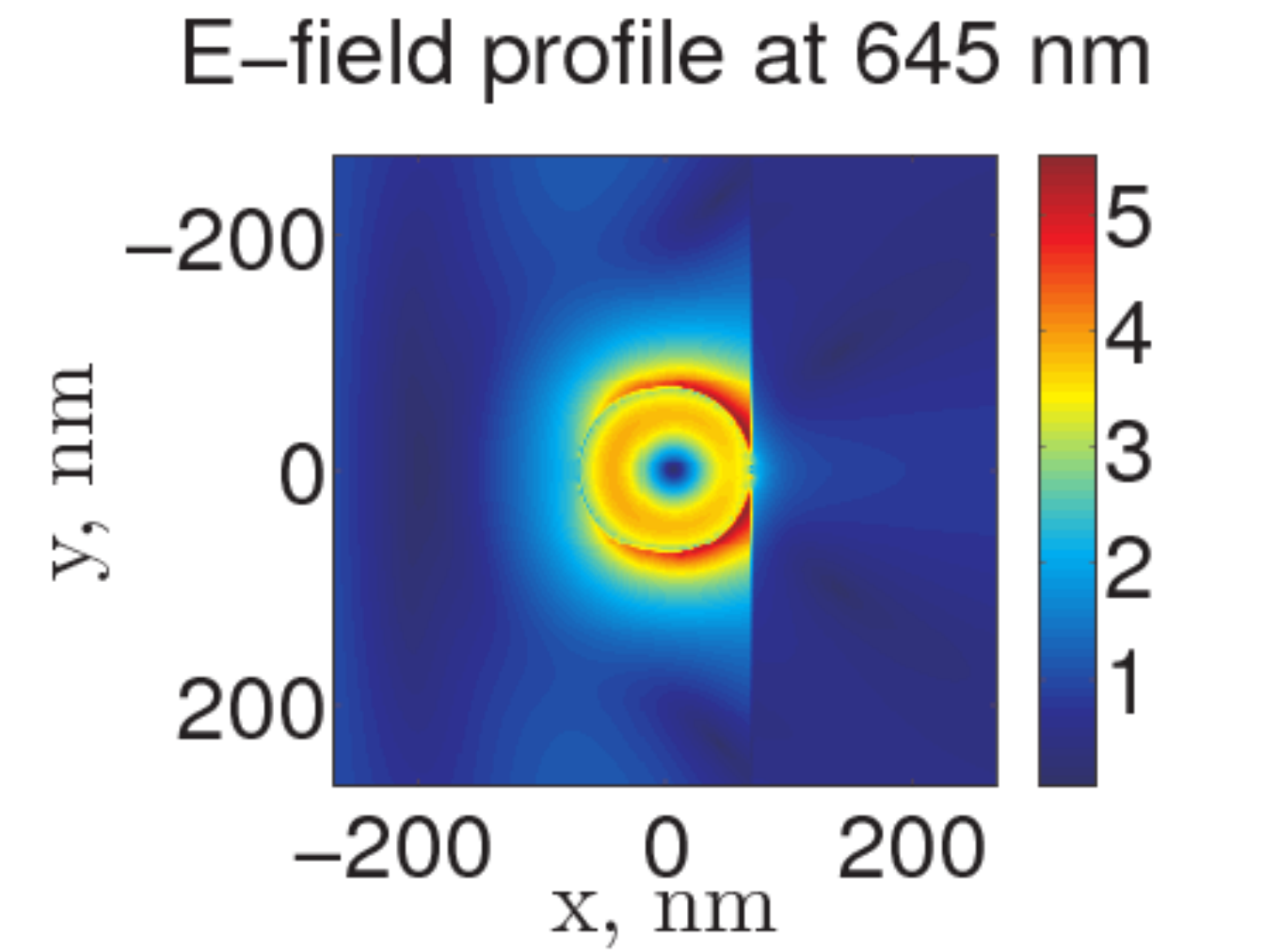}
    \end{center}
  \end{minipage}
  \vfill
  \caption{Optical properties of a dielectric nanoparticle
    ($\epsilon_{particle} = 20$) of 70 nm radius on a dielectric
    substrate ($\epsilon_{dielectric} = 3.1$): (a) scattering
    cross-section spectra, (b) the summary of the resonant
    wavelengths, the scattering cross-section normalised to the
    geometric cross-section, and electric $p_y$ and magnetic moments
    $m_z$, (c-e) the spatial distribution of the electric field
    amplitudes at resonant wavelengths (normalised to the incident
    field).}
  \label{dielectric_substrate}
\end{figure}

In many cases, dielectric particles are placed on glass substrates as
the result of their fabrication and for optical
characterisation~\cite{fu-directional-2013}. This type of substrates
is expected to introduce the smallest distortions in the optical
properties of a particle, compared to other types of substrates, e.g.,
metallic ones. We considered a glass substrate with $\epsilon$ = 3.1
corresponding to the family of flint glasses. Comparing to the
nanoparticles in free space, the glass substrate influence is in
significant suppression of the high-order multipoles
(Fig.~\ref{dielectric_substrate}), with both electric and, especially,
magnetic dipolar resonances being less affected. These conclusions are
confirmed by the field amplitude distributions at the resonant
wavelengths.  Minor electric field amplitude amplification for ED and
MD resonance and nearly two times reduction for MQ resonance can be
seen compared to the particle in free-space. The electric moment of
the particle is slightly increased while the magnetic one is slightly
reduced.

\clearpage
\subsection{PEC substrate}
\begin{figure}[h!]
  \begin{minipage}[h]{0.49\textwidth} a) \end{minipage} \hfill
  \begin{minipage}[h]{0.49\textwidth} b) \end{minipage} \vfill
  \begin{minipage}[h]{0.49\textwidth}
    \begin{center}
      \includegraphics[width=0.99\textwidth]{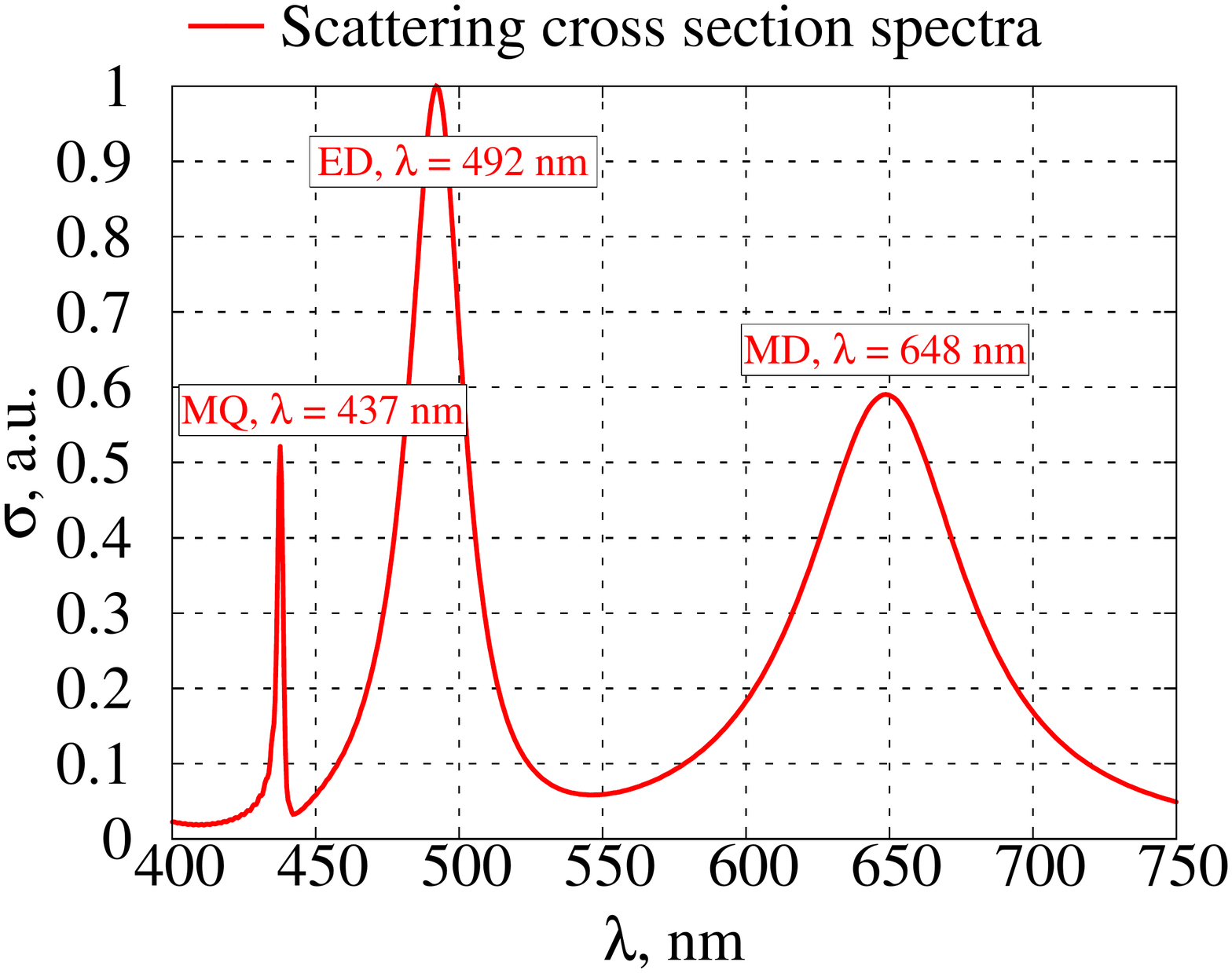}
    \end{center}
  \end{minipage}
  \hfill
  \begin{minipage}[h]{0.49\textwidth}
    \begin{center}
      {\normalfont
        \begin{tabular}{cccc}
          \toprule
          \multicolumn{4}{c}{Particle on a PEC substrate} \\
          \midrule
          $\lambda$, nm & 437 & 492 & 648\\
          \midrule
          $\sigma$ / S &  5.2 & 10.0 & 6.0\\
          \midrule
          $p_y$, [e $\cdot$ nm] &  2.4$e^{-2}$ & 7.2$e^{-2}$ &  7.0$e^{-3}$\\
          \midrule
          $m_z$, [A $\cdot$ nm$^2$] & 1.8 & 1.7 & 0.7\\
          \bottomrule
        \end{tabular}
      } \end{center}
  \end{minipage}
  \vfill
  \begin{minipage}[h]{0.3\textwidth} c) \end{minipage} \hfill
  \begin{minipage}[h]{0.3\textwidth} d) \end{minipage} \hfill
  \begin{minipage}[h]{0.3\textwidth} e) \end{minipage} \vfill
  \begin{minipage}[h]{0.3\textwidth}
    \begin{center}
      \includegraphics[width=0.99\textwidth]
                      {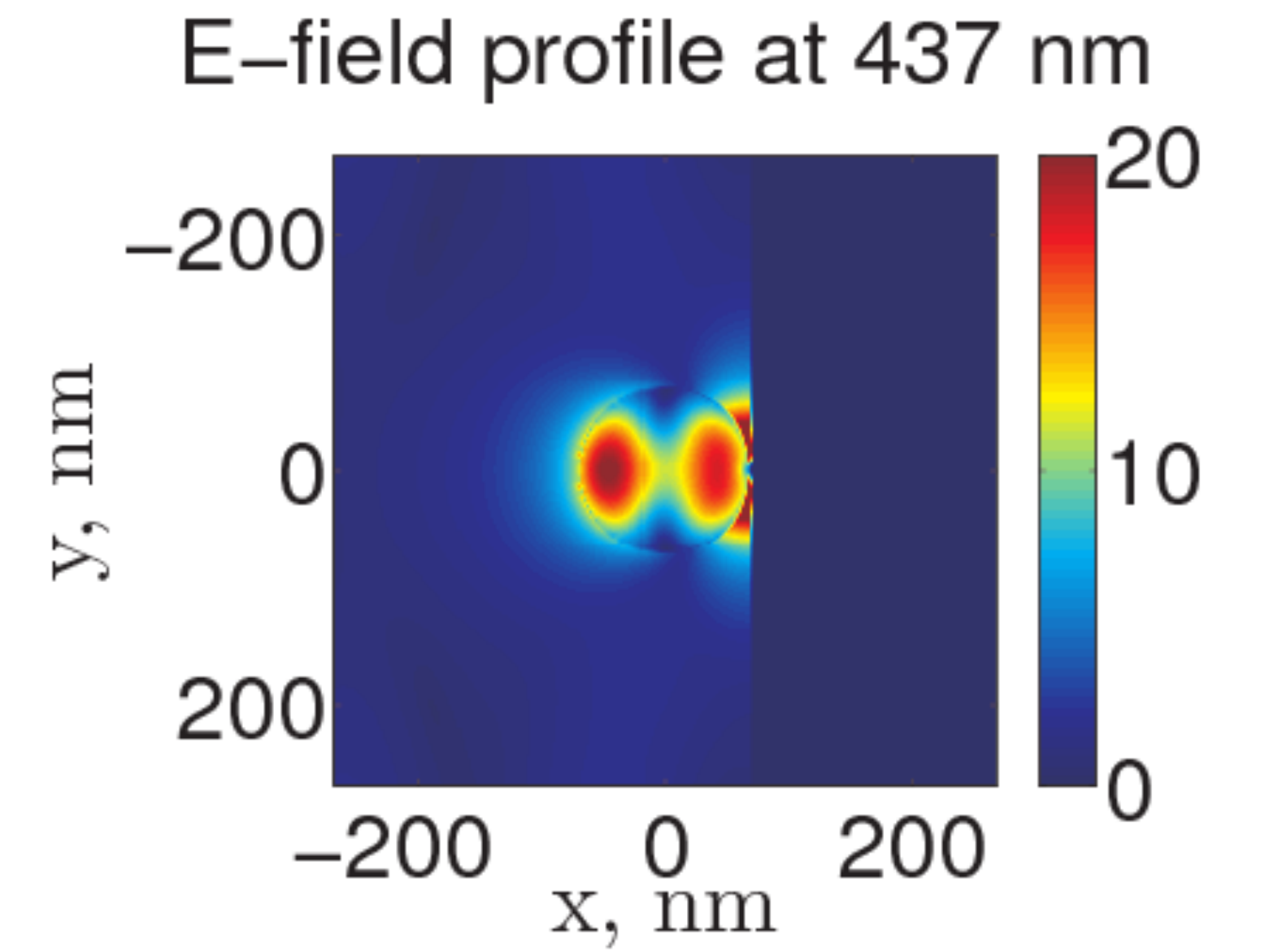}
    \end{center}
  \end{minipage}
  \hfill
  \begin{minipage}[h]{0.3\textwidth}
    \begin{center}
      \includegraphics[width=0.99\textwidth]
                      {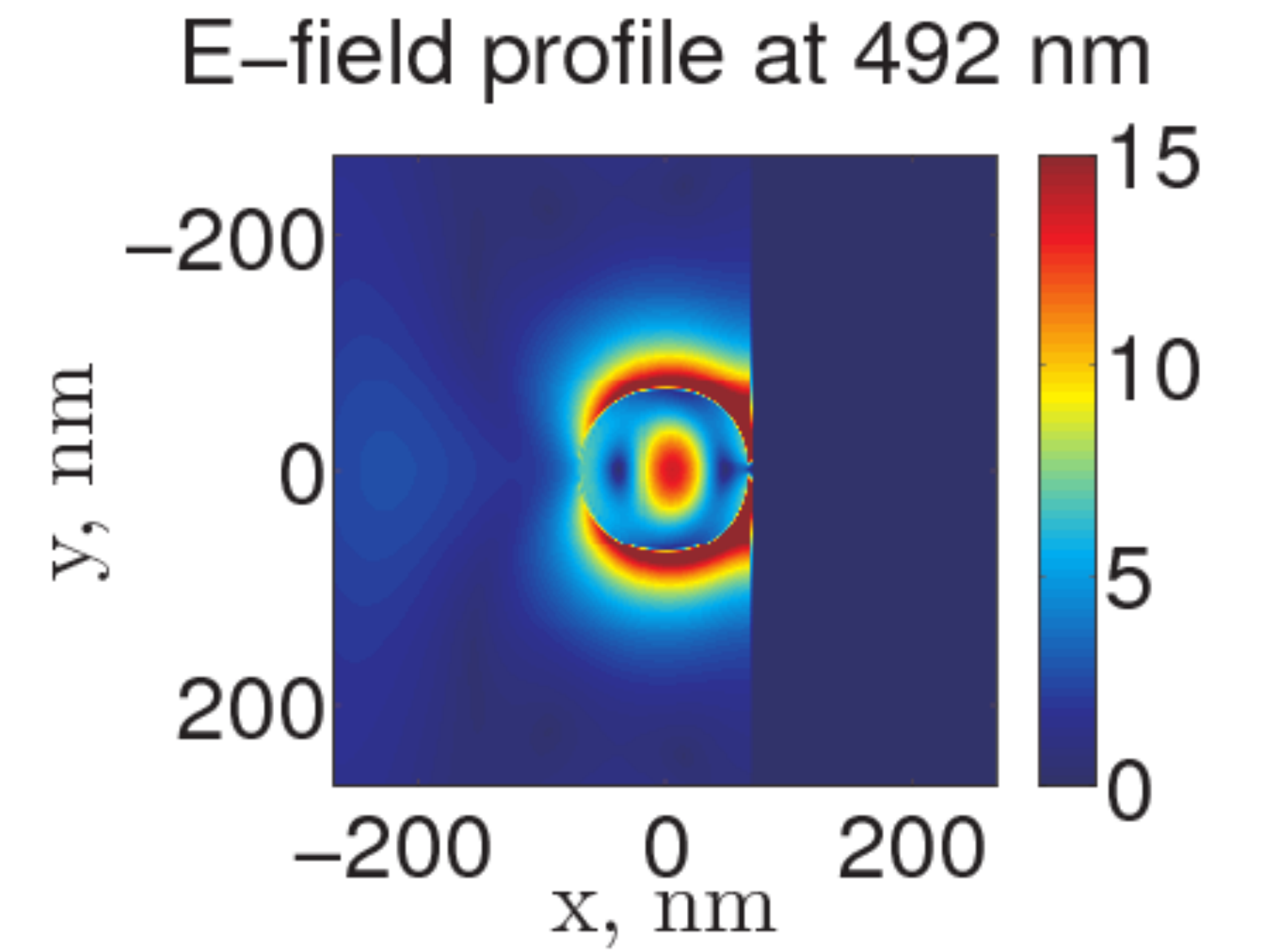}
    \end{center}
  \end{minipage}
  \hfill
  \begin{minipage}[h]{0.3\textwidth}
    \begin{center}
      \includegraphics[width=0.99\textwidth]
                      {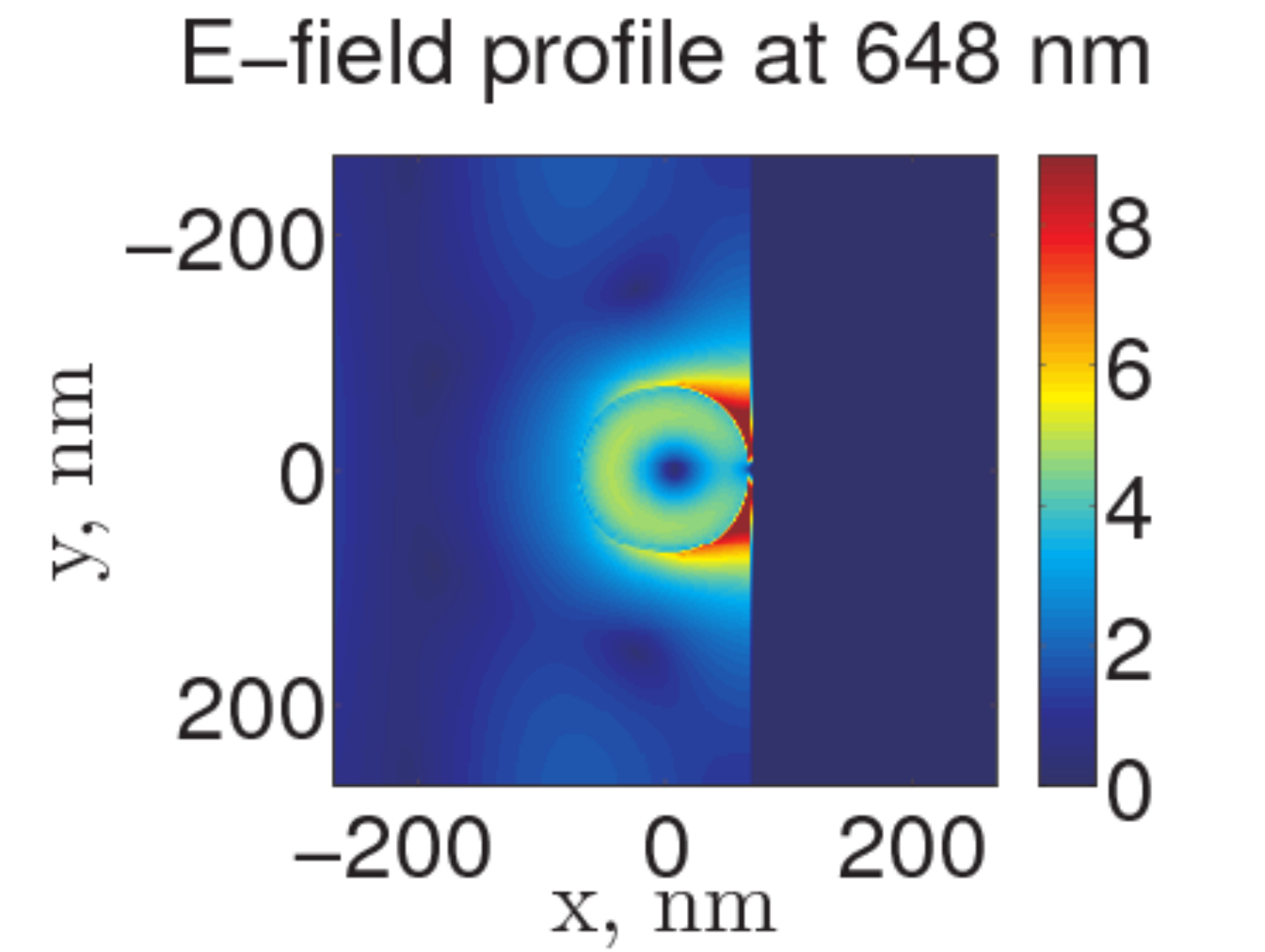}
    \end{center}
  \end{minipage}
    \caption{Optical properties of a dielectric nanoparticle
      ($\epsilon_{particle} = 20$) of 70 nm radius on a PEC substrate
      ($\epsilon_{PEC} = 1 + 1e^6i$): (a) scattering cross-section
      spectra, (b) the summary of the resonant wavelengths, the
      scattering cross-section normalised to the geometric
      cross-section, and electric $p_y$ and magnetic moments $m_z$,
      (c-e) the spatial distribution of the electric field amplitudes
      at resonant wavelengths (normalised to the incident field).}
  \label{pec_substrate}
\end{figure}

In order to test the applicability of the image theory for high-index
dielectric particles, a perfect electric conductor (PEC) substrate was
considered. This dispersionless metal with infinitely large imaginary
part of the permittivity acts like a perfect mirror for both
electromagnetic waves and point charges. The results for a PEC
substrate are summarized in Fig.~\ref{pec_substrate}. Compared the
free space, a PEC substrate leads to the spectral shift of the
nanoparticle resonances and significant modifications of the spectrum
and magnitude of the scattering cross-sections. For example, the
electric dipole, induced in the direction parallel to the surface,
should create the contra-oriented image dipole and, as the result, the
resulting dark quadrupolar resonance of the system should suppress the
scattering (as in the case of subwavelength plasmonic
particles). Nevertheless, the electric dipole resonance became the
strongest in the presence of PEC. This effect is related to the
retardation, since the optical path between the dielectric sphere and
its image become comparable to wavelength due to the high-index
dielectric. Near-field amplitudes are almost three-fold increased
compared to free space as the result of perfect reflection of the
incident wave by the substrate. Both electric and magnetic moments in
this case experience strong resonant amplification by almost two
orders of magnitude.

\clearpage
\subsection{Gold substrate}
\begin{figure}[h!]
  \begin{minipage}[h]{0.49\textwidth} a) \end{minipage} \hfill
  \begin{minipage}[h]{0.49\textwidth} b) \end{minipage} \vfill
  \begin{minipage}[h]{0.49\textwidth}
    \begin{center}
      \includegraphics[width=0.99\textwidth]{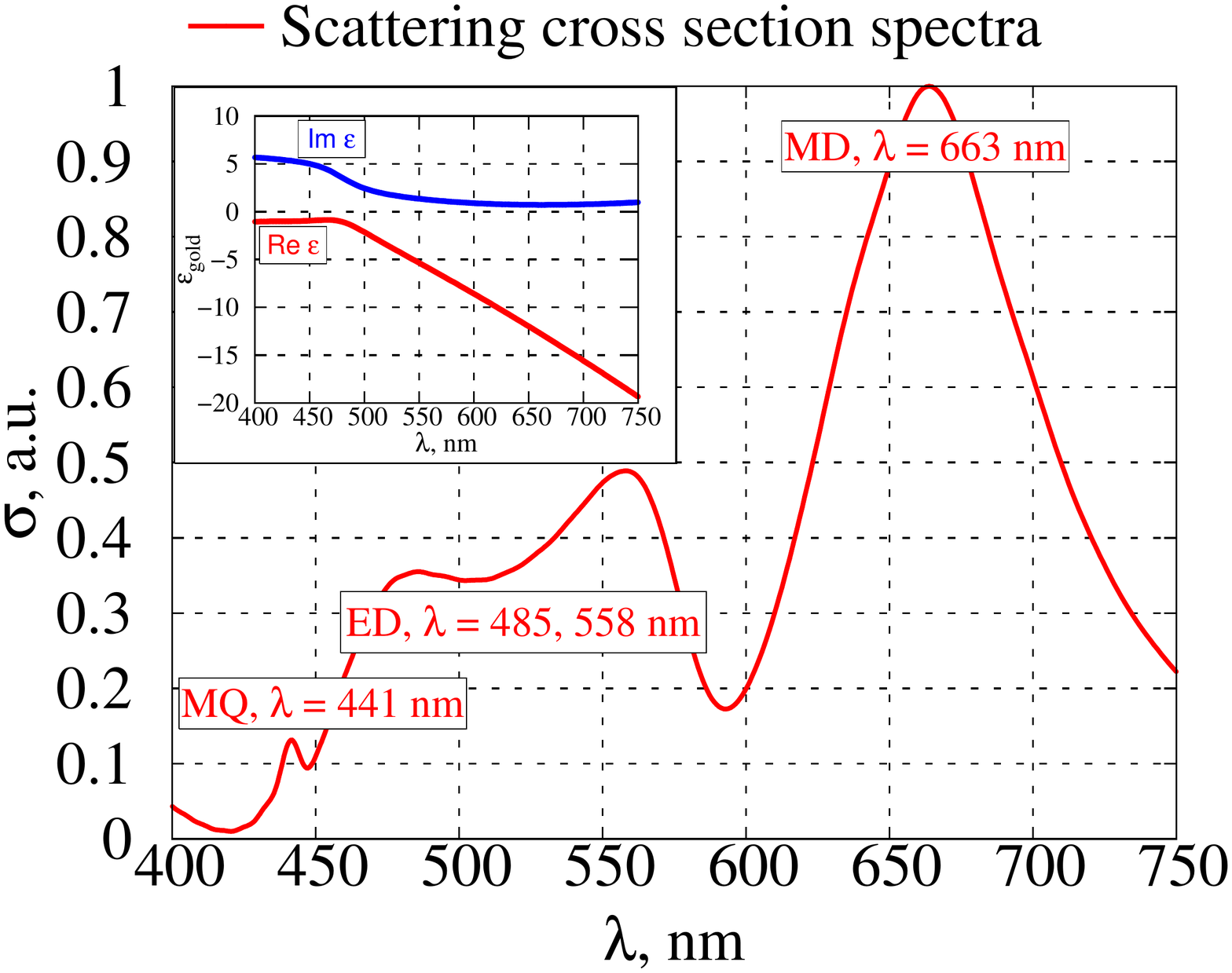}
    \end{center}
  \end{minipage}
  \hfill
  \begin{minipage}[h]{0.49\textwidth}
    \begin{center}
      {\normalfont
        \begin{tabular}{ccccc}
          \toprule
          \multicolumn{5}{c}{Particle on a Au substrate} \\
          \midrule
          $\lambda$, nm & 441 & 485 & 558 & 663\\
          \midrule
          $\sigma$ / S & 0.5 & 1.4 & 1.9 & 3.9\\
          \midrule
          $p_y$, [e $\cdot$ nm] & 7.0e$^{-5}$ & 1.1$e^{-3}$ & 2.3$e^{-2}$ & 1.7$e^{-3}$\\
          \midrule
          $m_z$, [A $\cdot$ nm$^2$] & 1.6$e^{-3}$ & 1.5$e^{-3}$ & 5.2$e^{-3}$ & 1.2$e^{-2}$\\
          \bottomrule
        \end{tabular}
      } \end{center}
  \end{minipage}
  \vfill
  \begin{minipage}[h]{0.24\textwidth} c) \end{minipage} \hfill
  \begin{minipage}[h]{0.24\textwidth} d) \end{minipage} \hfill
  \begin{minipage}[h]{0.24\textwidth} e) \end{minipage} \hfill
  \begin{minipage}[h]{0.24\textwidth} f) \end{minipage} \vfill
  \begin{minipage}[h]{0.24\textwidth}
    \begin{center}
      \includegraphics[width=0.99\textwidth]
      {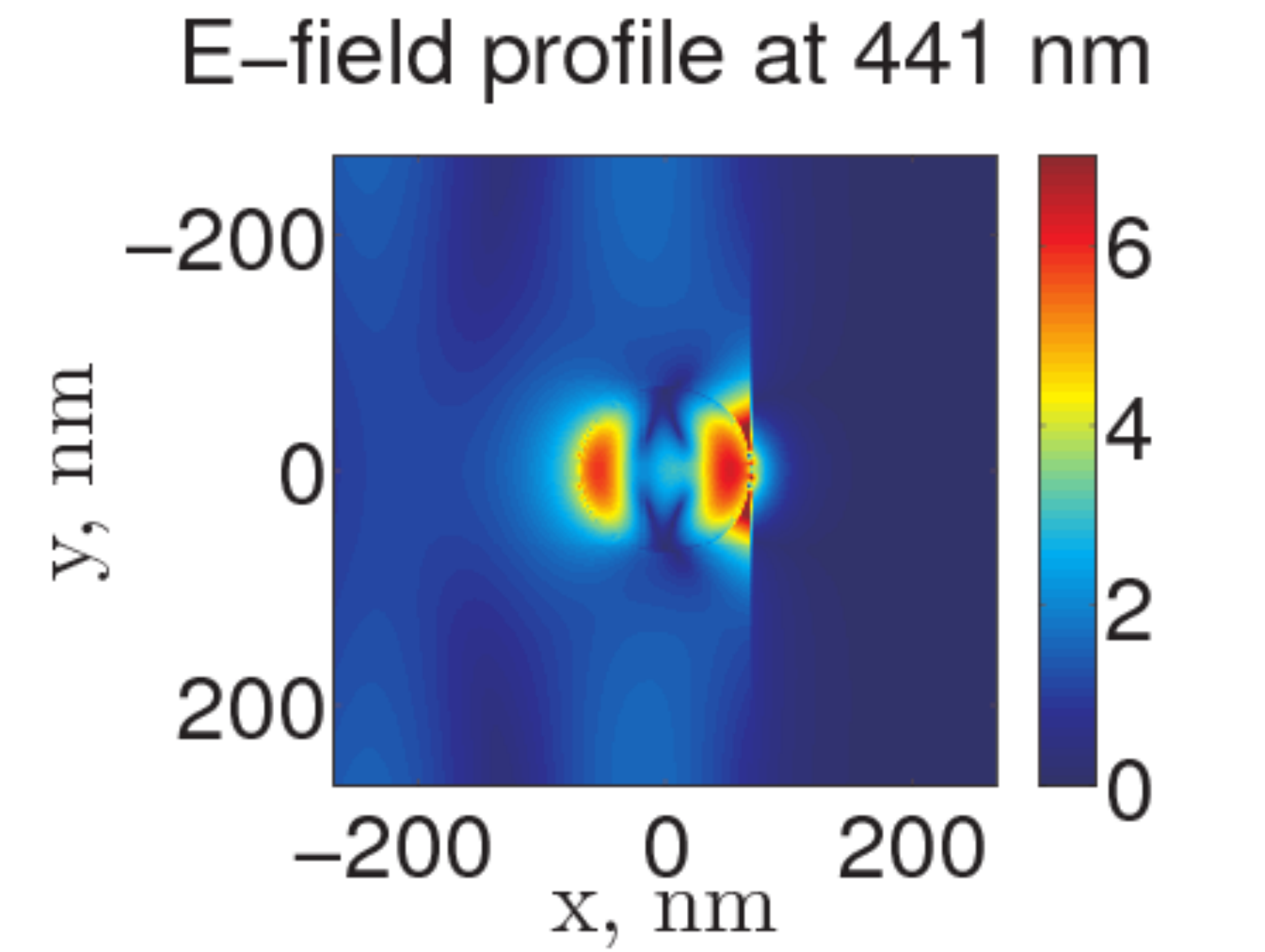}
    \end{center}
  \end{minipage}
  \hfill
  \begin{minipage}[h]{0.24\textwidth}
    \begin{center}
      \includegraphics[width=0.99\textwidth]
      {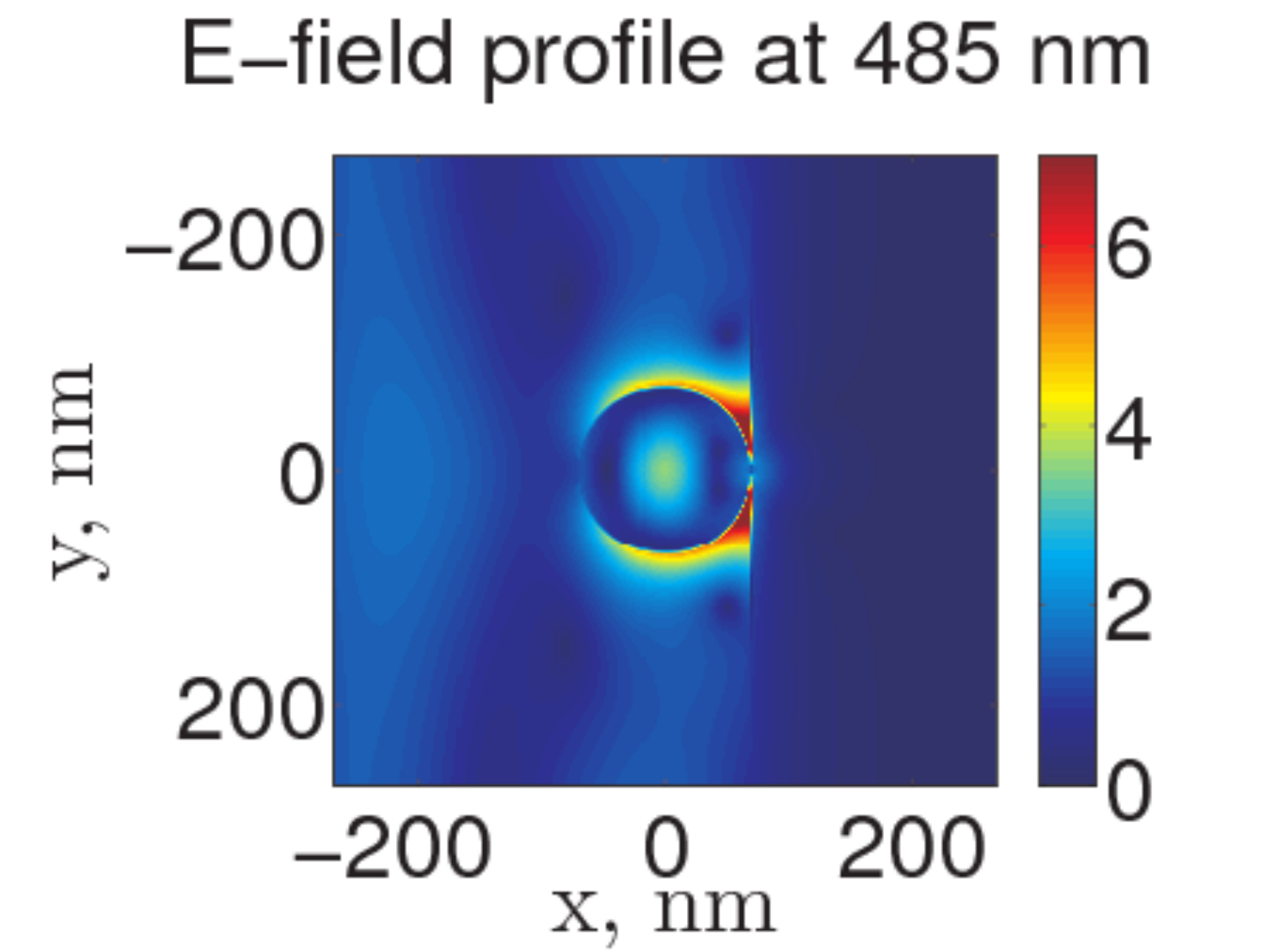}
    \end{center}
  \end{minipage}
  \hfill
  \begin{minipage}[h]{0.24\textwidth}
    \begin{center}
      \includegraphics[width=0.99\textwidth]
      {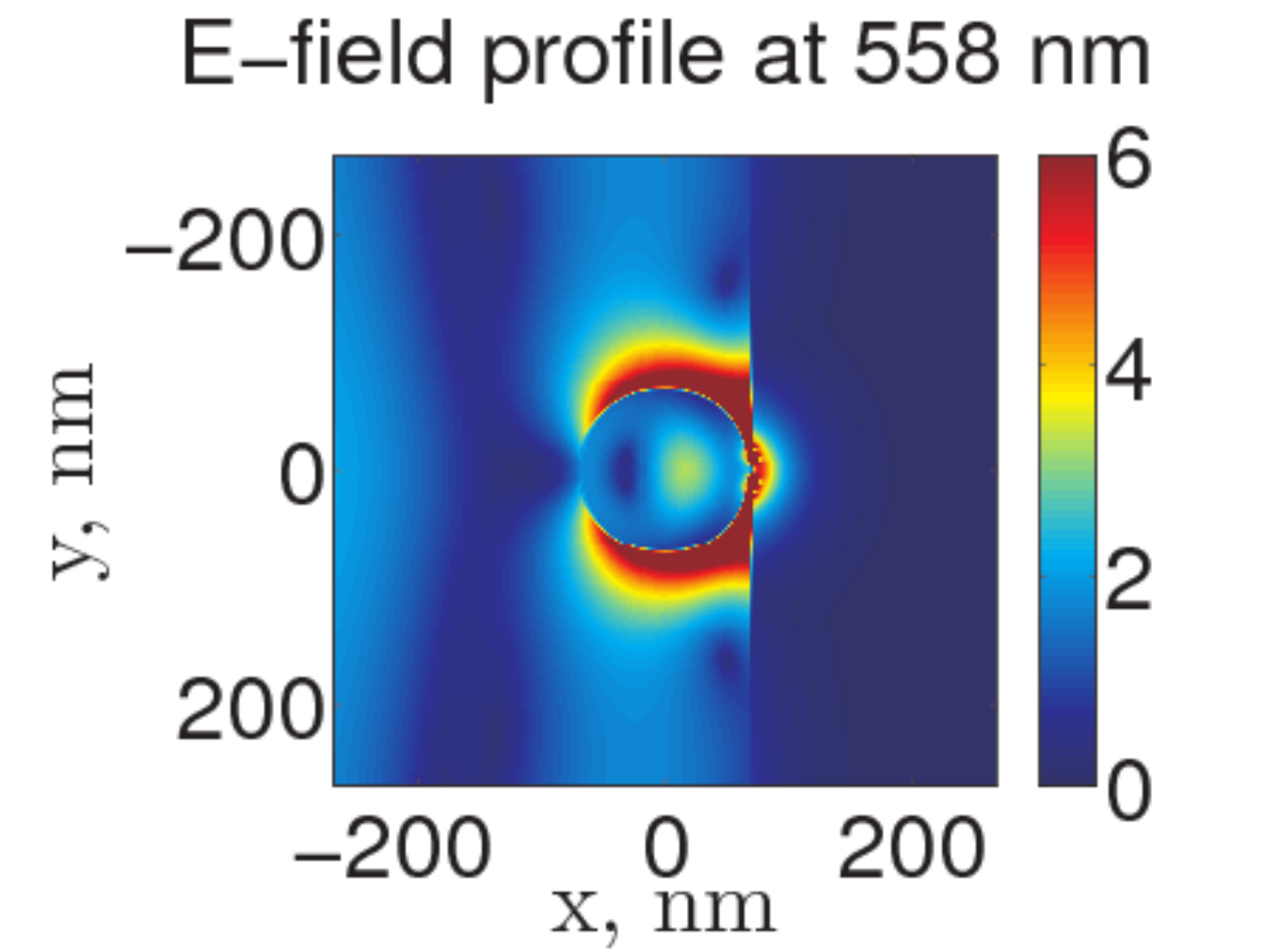}
    \end{center}
  \end{minipage}
  \hfill
  \begin{minipage}[h]{0.24\textwidth}
    \begin{center}
      \includegraphics[width=0.99\textwidth]
      {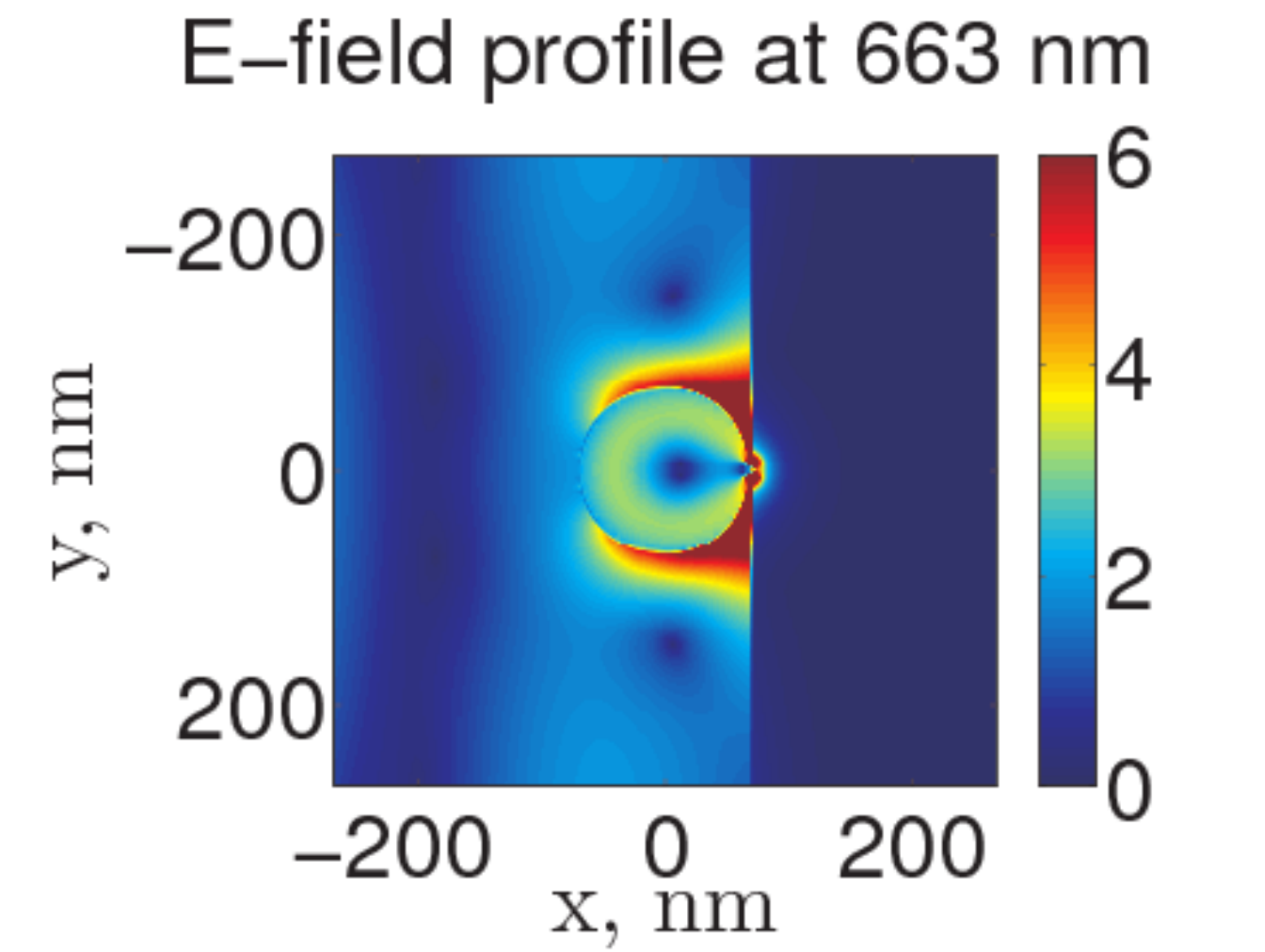}
    \end{center}
  \end{minipage}
      \caption{Optical properties of a dielectric nanoparticle
        ($\epsilon_{particle} = 20$) of 70 nm radius on a gold
        substrate (inset if (a) shows $\epsilon_{gold}$ taken
        from~\cite{johnson-optical-1972}): (a) scattering
        cross-section spectra, (b) the summary of the resonant
        wavelengths, the scattering cross-section normalised to the
        geometric cross-section, and electric $p_y$ and magnetic
        moments $m_z$, (c-e) the spatial distribution of the electric
        field amplitudes at resonant wavelengths (normalised to the
        incident field). }
  \label{gold_substrate}
\end{figure}

The illumination of a scatterer placed on a gold film gives rise to
excitation of surface plasmon
polaritons~\cite{maier-plasmonics-2007}. The excitation of additional
propagating surface wave substantially changes the optical properties
of dielectric spheres (Fig.~\ref{gold_substrate}). It can be seen that
the electric dipole resonance is split in two. This is the result of
strong coupling by anti-crossing between two coupled dipoles (SPP mode
and ED resonance of the particle).  Moreover, the electric dipole mode
excites the SPP with much higher efficiency than the other particle
resonances. Electric field amplitudes experience minor amplification
comparing to the case of free space, except for the MQ resonance. The
electric moment experiences nearly ten fold enhancement due to the
presence of the substrate, whilst the magnetic moments retain the
values close to those in free space.

\clearpage
\subsection{Metal-dielectric multilayered substrate:
  comparison between effective medium and composite representations}
Hyperbolic metamaterials are artificially created media with strongly
anisotropic effective permittivity tensors. They attract considerable
attention due to their potential to substantially increase the local
density of states and, as the result, increase radiative rate of
emitters situated in their vicinity. Here, we consider the impact of a
hyperbolic material substrates on the optical properties of high-index
dielectric particles. One of the yet open questions is the
applicability of effective medium theory to various physical scenarios
considering an emitter (or scatterer) placed in the near-field
proximity to a metamaterial~\cite{ginzburg-self-induced-2013,
  kapitanova-photonic-2014}. Hereafter, we compare the layered
realization of hyperbolic metamaterial~\cite{poddubny-hyperbolic-2013}
with its homogeneous counterpart described via effective medium
theory~\cite{fainberg-artificial-1955} neglecting the effects of
spatial dispersion~\cite{orlov-strong-2010}.

\begin{figure}[h!]
  \begin{minipage}[h]{0.45\textwidth} a)  \end{minipage} \hfill
  \begin{minipage}[h]{0.45\textwidth} b)  \end{minipage} \vfill
  \begin{minipage}[h]{0.49\textwidth}
    \begin{center}
      \includegraphics[width=0.99\textwidth]{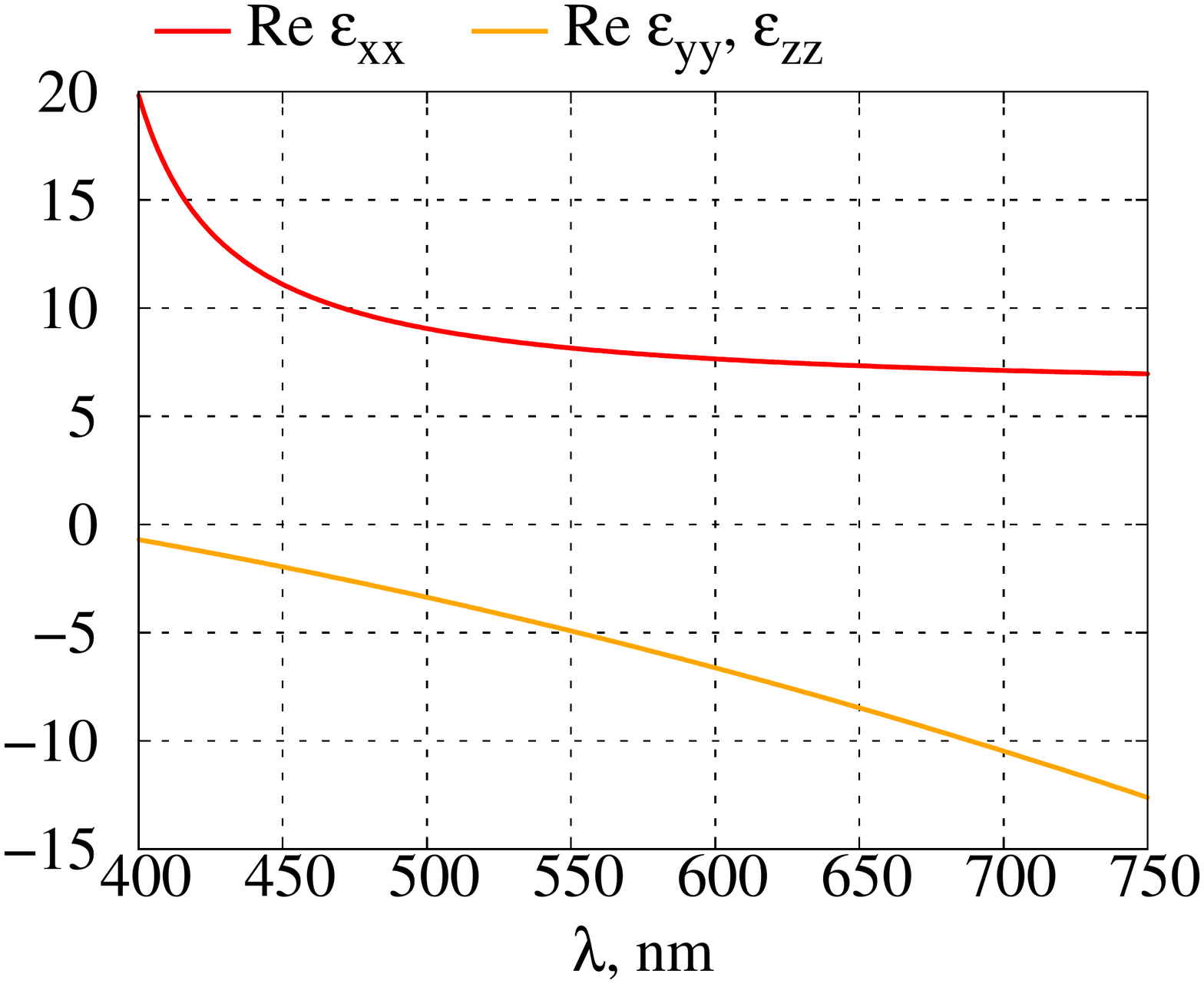}
    \end{center}
  \end{minipage}
  \hfill
  \begin{minipage}[h]{0.49\textwidth}
    \begin{center}
      \includegraphics[width=0.99\textwidth]{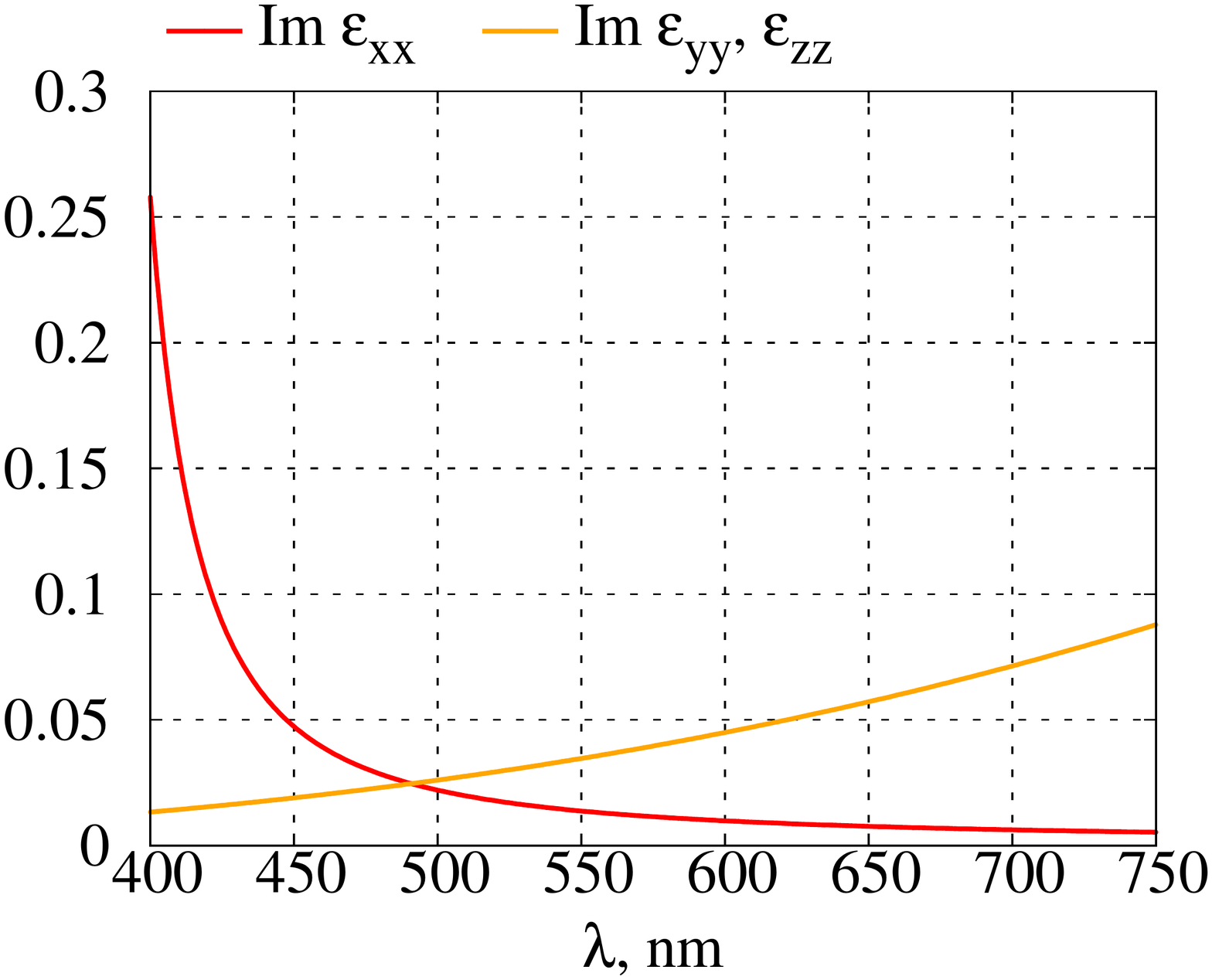}
    \end{center}
  \end{minipage}
  \caption{Dispersion of the components of the effective permittivity
    $\epsilon_{xx}, \epsilon_{yy} = \epsilon_{zz}$, for the
    metamaterial substrate made of a silver-glass multilayer: (a)real
    and (b) imaginary parts.}
  \label{material_dispersion}
\end{figure}

The considered nanostructured substrate consists of dielectric layers
with $\epsilon_{d} = 3.1$ and silver metal layers with $\epsilon_m$
(the Drude model for silver for
considered~\cite{johnson-optical-1972}). The layers have the same
subwavelength thickness (30 nm) and can be described using a
diagonalised effective permittivity tensor with components
\begin{eqnarray}
  \epsilon_{xx} = \frac{\epsilon_{d} \cdot \epsilon_{m}}{(1 - \rho) \cdot \epsilon_{d} + \rho \cdot \epsilon_{m}} \nonumber \\
  \epsilon_{yy} = \epsilon_{zz} = \rho \cdot \epsilon_{d} + (1 - \rho) \cdot \epsilon_{m},
\label{emt_calculation}
\end{eqnarray}
where $\rho$=0.5 is the filling factor in the case of equal metal and
dielectric layer thicknesses. These effective permittivity components
are shown in Fig.~\ref{material_dispersion}.

The scattering cross-sections for the particle placed on the
substrates made of the multilayered composite and the effective medium
look very much alike. At the same time, the field distributions
corresponding to the excited resonances are substantially different
depending on the choice of the substrate. Clearly distinguishable
radiation cones (shaped as horizontal ``V'' with varying opening
angles) are visible in the effective medium substrate, but they are
suppressed in metal-dielectric slab (cf. Fig. 4 in
Ref.~\cite{ginzburg-self-induced-2013}). This is the result of the
near-field interaction of the nanoparticle with the first metal layer,
leading to the breakdown of the effective medium approach valid for
plane waves. Measuring the values of radiation opening angles at the
resonant wavelengths in the far-field after passing through the
substrate gives the opportunity to determine the type of a
nanoparticle resonance.  The magnetic moment of the particle at the
magnetic dipole resonance is reduced on the metamaterial substrate in
both descriptions, and other moments are slightly increased.  The
presence of hyperbolic metamaterial reduces the reflection of the
incident wave from the substrate. The scattering particle has an
access to the high-density of states in the substrate, reducing the
backscattering efficiency. This complex interplay results in the
aforementioned changes of the values of moments. Interestingly, while
the field profiles inside the metamaterial substrate are remarkably
different in the layered and homogeneous realizations, the overall
system's responses in the far field above the substrate are similar.

\begin{figure}[h!]
\begin{center}
  \begin{minipage}[h]{0.99\textwidth}
    \begin{tabular}{cc}
      \toprule
      Metal-dielectric structure & Effective medium material\\
      \midrule
      a)~~~~~~~~~~~~~~~~~~~~~~~~~~~~~~~~~~~~~~~~ & b)~~~~~~~~~~~~~~~~~~~~~~~~~~~~~~~~~~~~~~~~\\
      \includegraphics[width=0.49\textwidth]{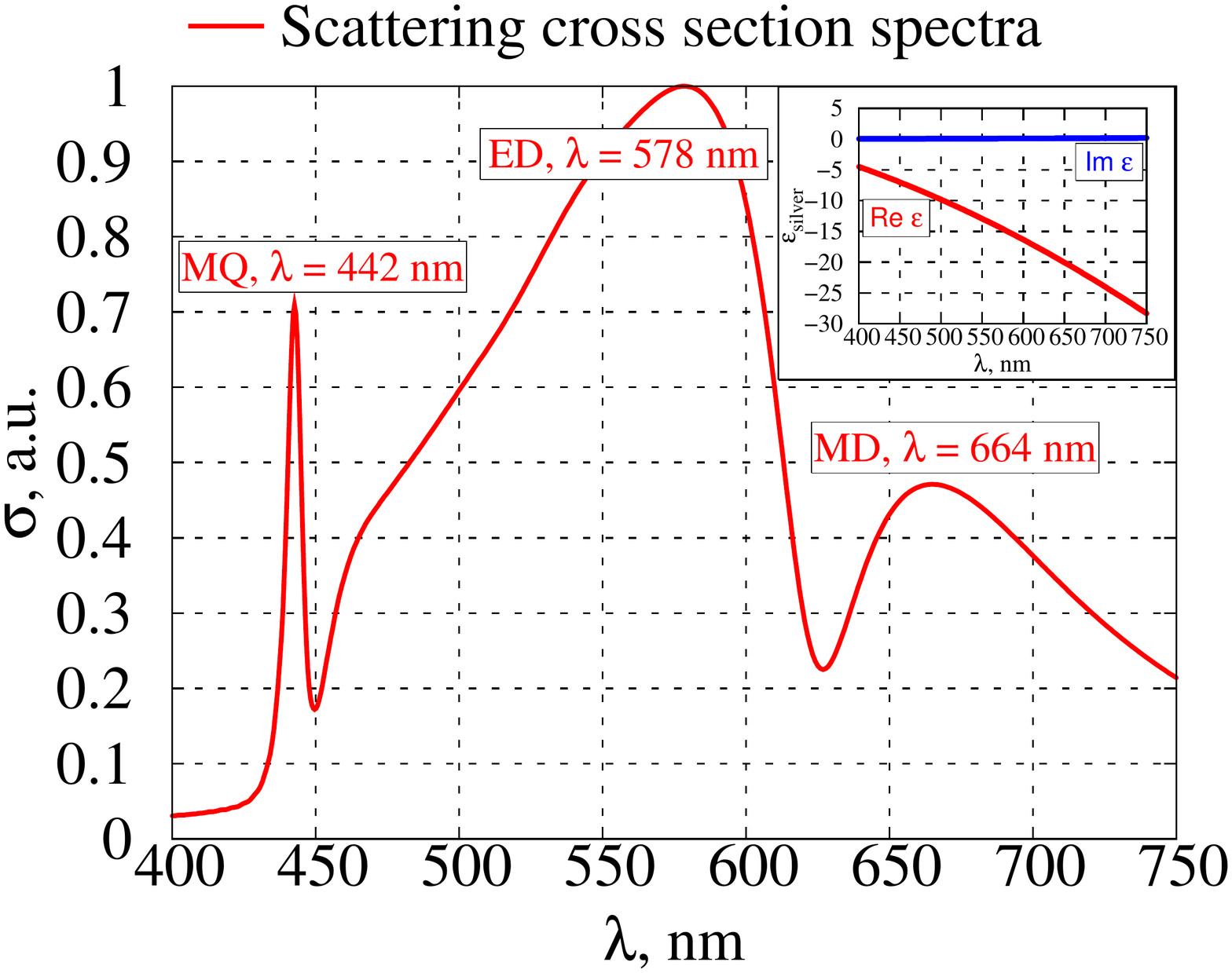}&
      \includegraphics[width=0.49\textwidth]{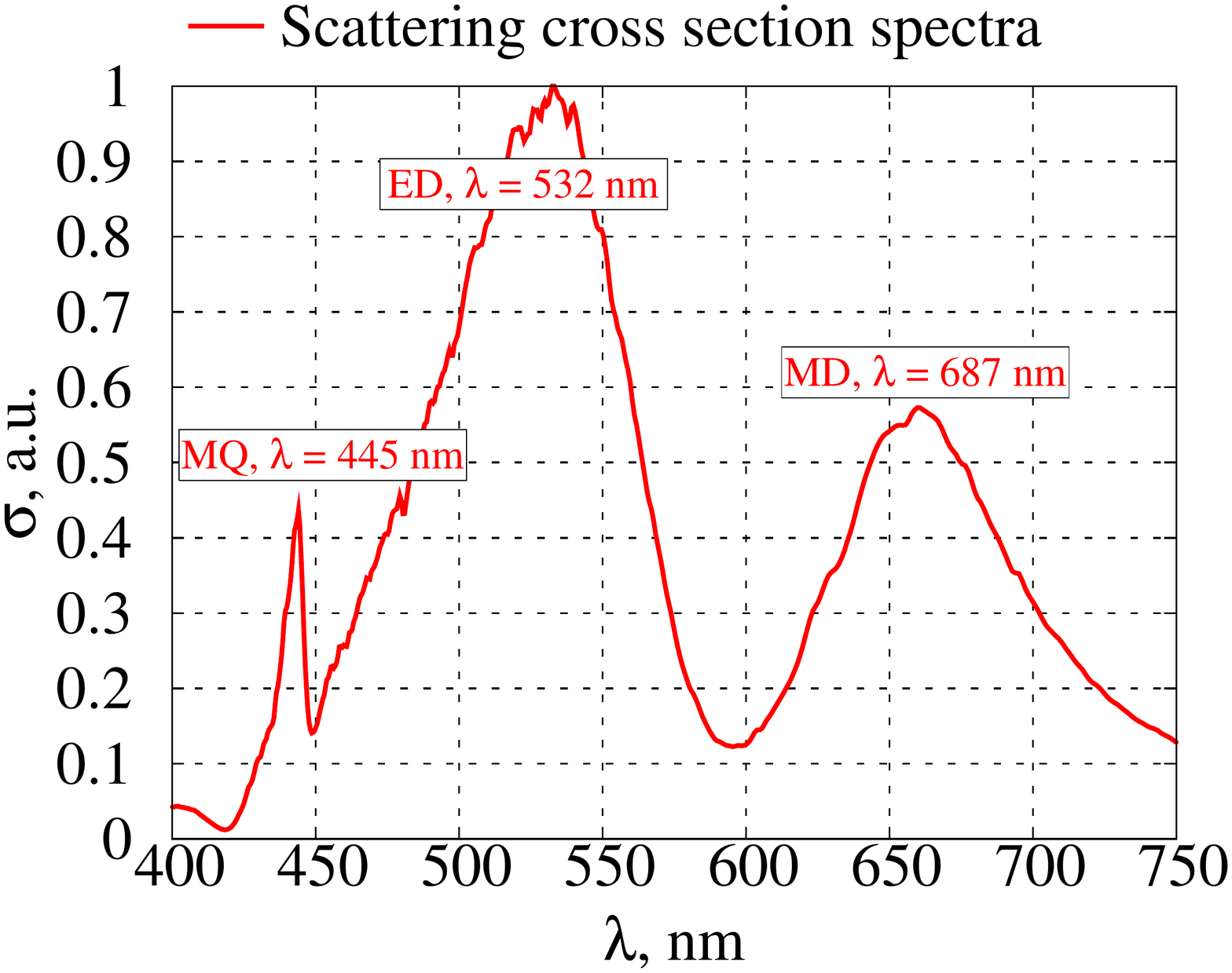}\\
      \midrule
      c)~~~~~~~~~~~~~~~~~~~~~~~~~~~~~~~~~~~~~~~~ & d)~~~~~~~~~~~~~~~~~~~~~~~~~~~~~~~~~~~~~~~~\\
      {\small
        \begin{tabular}{cccc}
          \midrule
          $\lambda$, nm & 442 & 578 & 664\\
          \midrule
          $\sigma$ / S & 3.1 & 4.4 & 2.0\\
          \midrule
          $p_y$, [e $\cdot$ nm] & 1.4e${^{-4}}$ & 2.2$e^{-3}$ & 1.6$e^{-3}$\\
          \midrule
          $m_z$, [A $\cdot$ nm$^2$] & 1.9$e^{-3}$ & 6.4$e^{3}$ & 6.6$e^{-3}$\\
        \end{tabular}
      } &
      {\small
        \begin{tabular}{cccc}
          \midrule
          $\lambda$, nm & 445 & 532 & 687\\
          \midrule
          $\sigma$ / S & 2.2 & 6.3 & 3.6\\
          \midrule
          $p_y$, [e $\cdot$ nm] & 2.0e$^{-4}$ & 2.8$e^{-3}$ & 1.7$e^{-3}$\\
          \midrule
          $m_z$, [A $\cdot$ nm$^2$] & 2.1$e^{-3}$ & 5.1$e^{-3}$ & 9.2$e^{-3}$\\
        \end{tabular}
      }\\
      \midrule
      e)~~~~~~~~~~~~~~~~~~~~~~~~~~~~~~~~~~~~~~~~ & f)~~~~~~~~~~~~~~~~~~~~~~~~~~~~~~~~~~~~~~~~\\
      \includegraphics[width=0.3\textwidth]{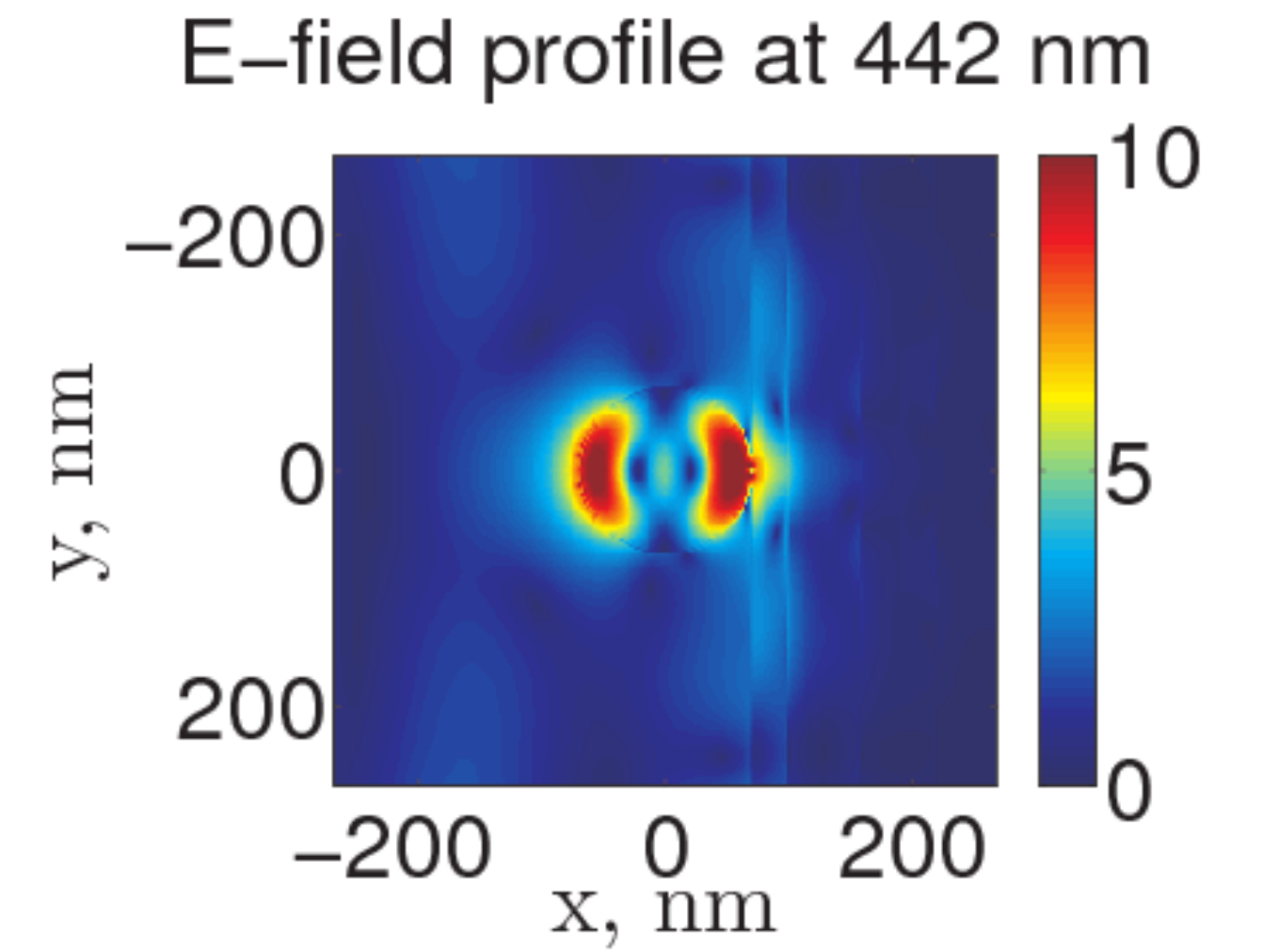} &
      \includegraphics[width=0.3\textwidth]{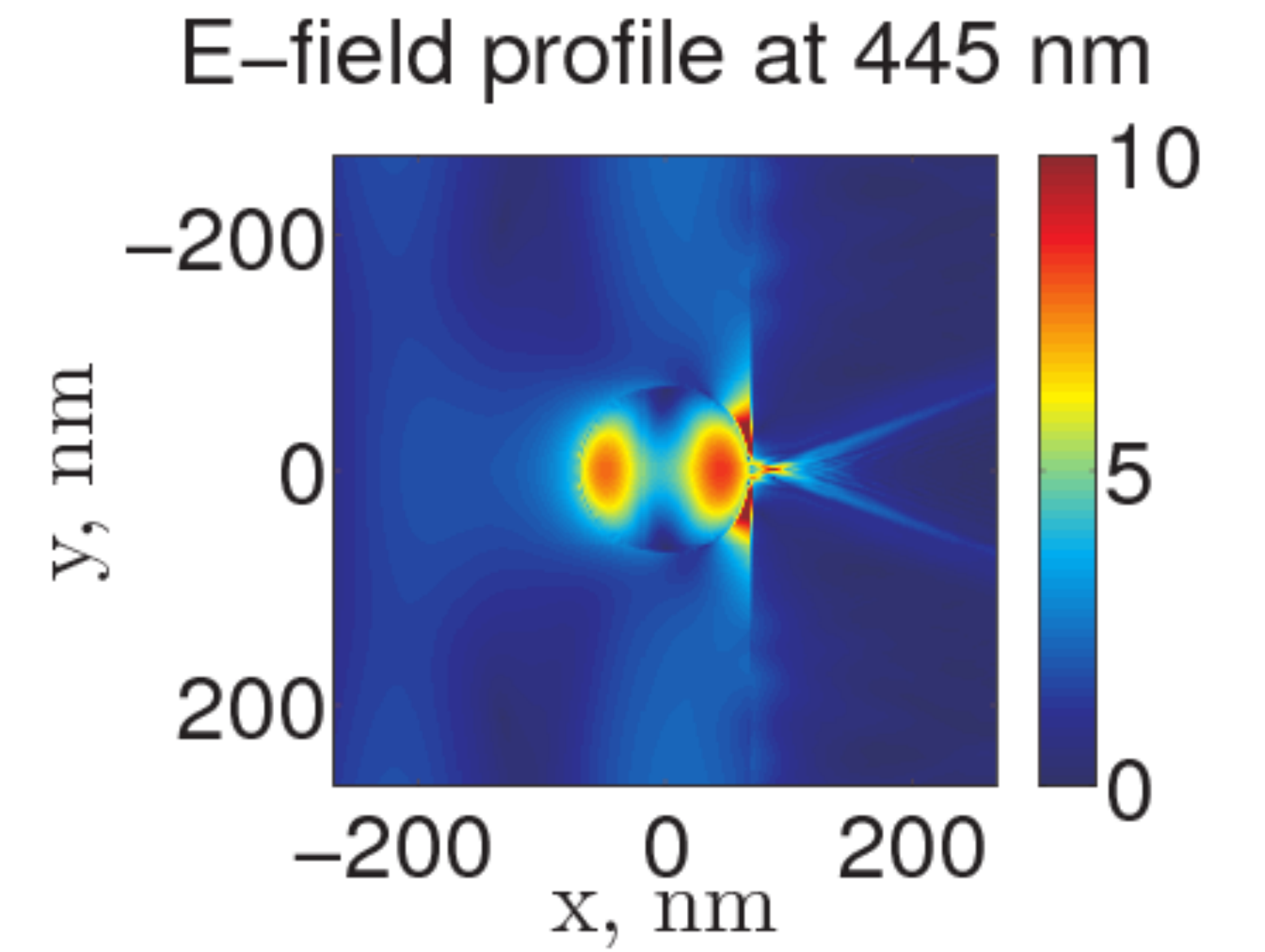} \\
      \midrule
      g)~~~~~~~~~~~~~~~~~~~~~~~~~~~~~~~~~~~~~~~~ & h)~~~~~~~~~~~~~~~~~~~~~~~~~~~~~~~~~~~~~~~~\\
      \includegraphics[width=0.3\textwidth]{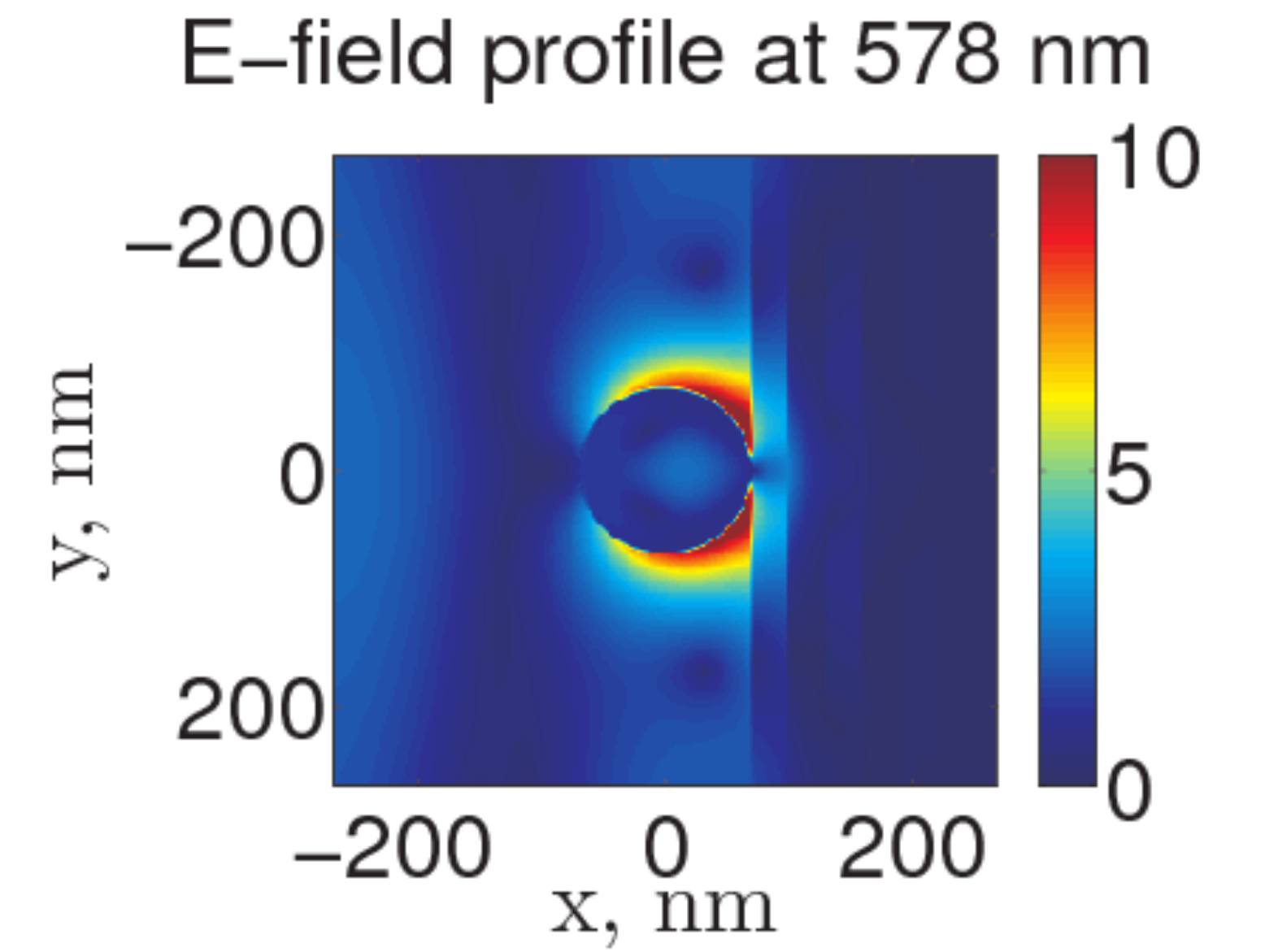} &
      \includegraphics[width=0.3\textwidth]{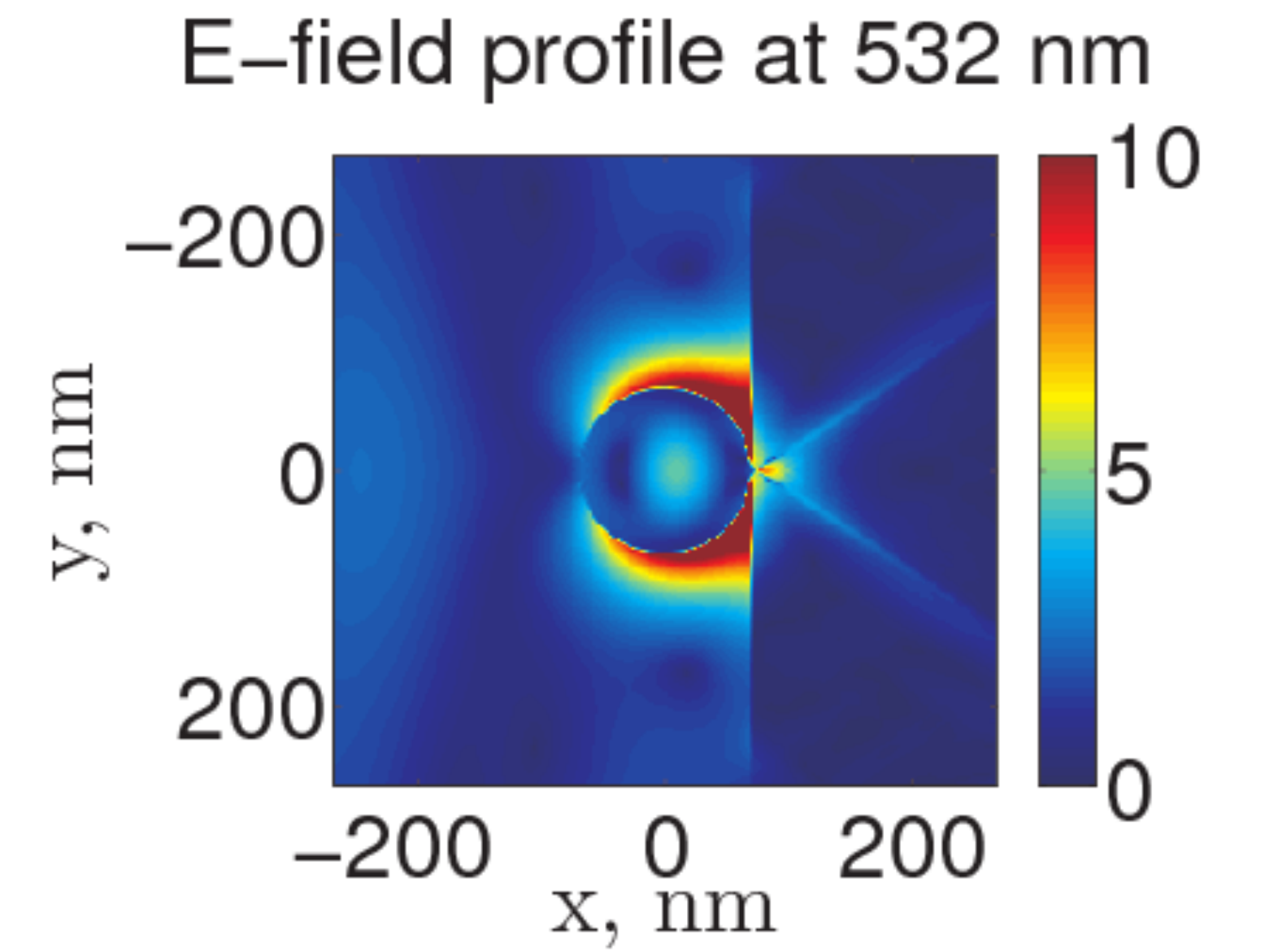} \\
      \midrule
      i)~~~~~~~~~~~~~~~~~~~~~~~~~~~~~~~~~~~~~~~~ & j)~~~~~~~~~~~~~~~~~~~~~~~~~~~~~~~~~~~~~~~~\\
      \includegraphics[width=0.3\textwidth]{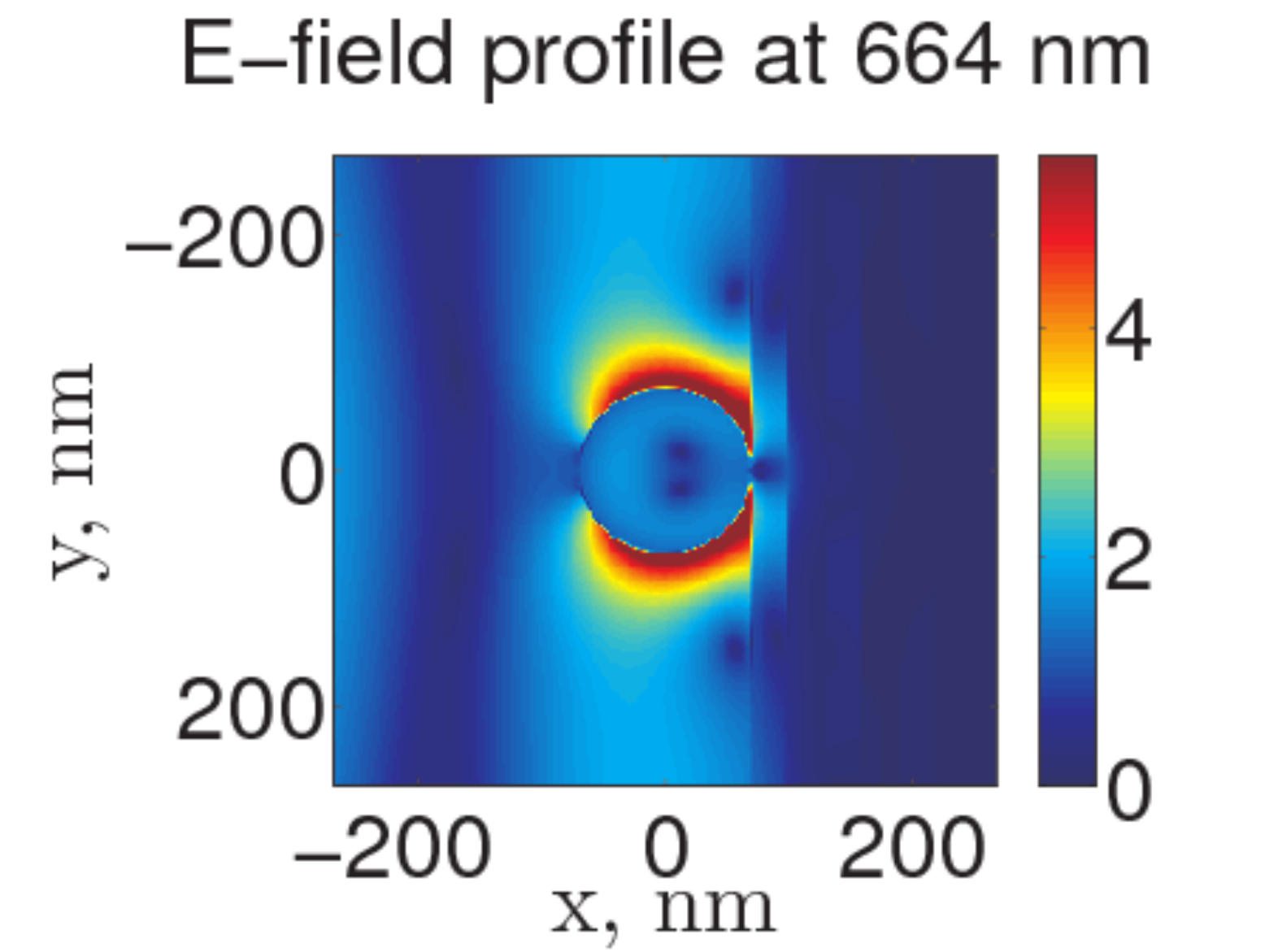} &
      \includegraphics[width=0.3\textwidth]{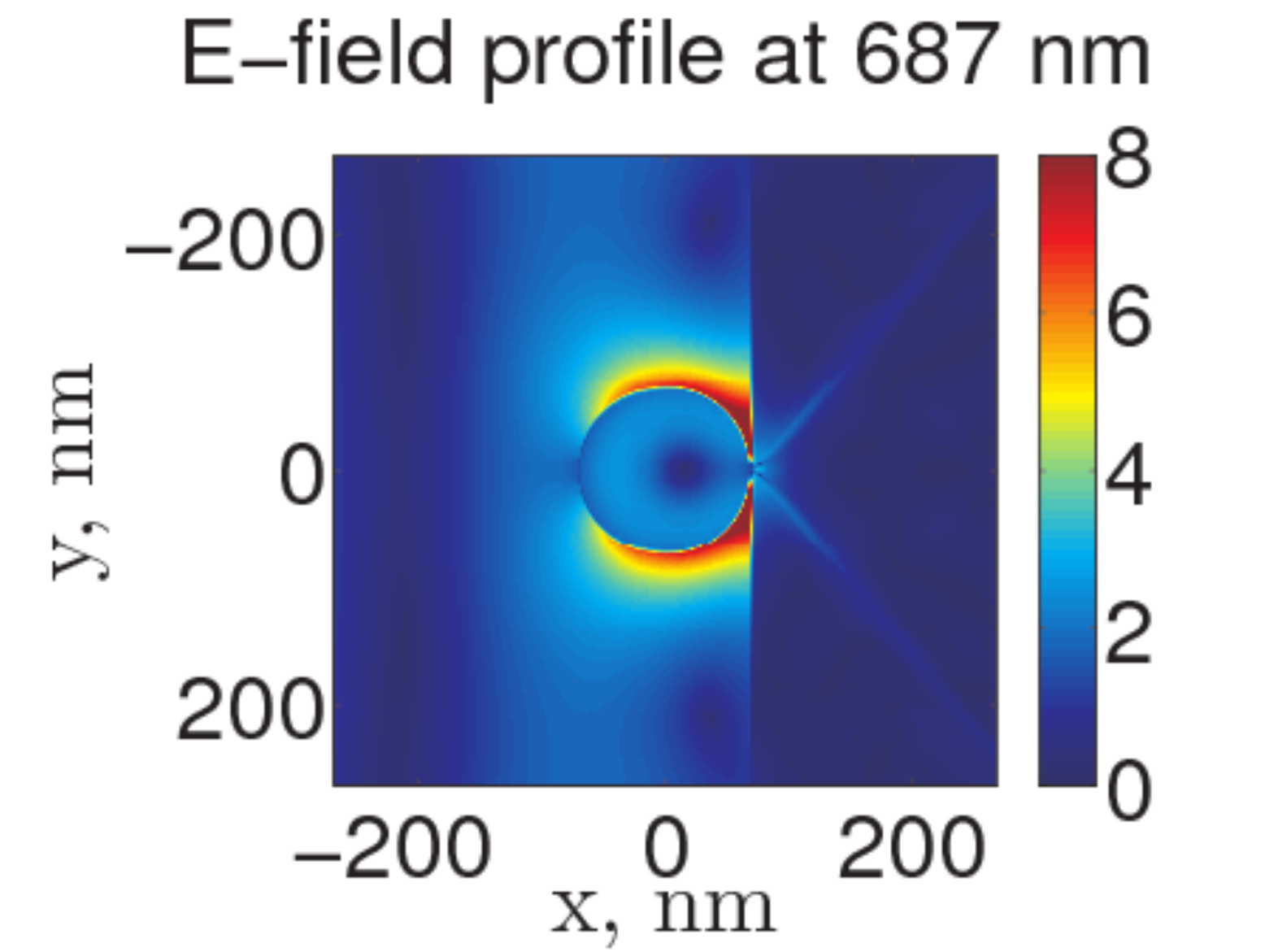} \\
      \bottomrule
    \end{tabular}
  \end{minipage}
      \caption{Optical properties of a dielectric nanoparticle
        ($\epsilon_{particle} = 20$) of 70 nm radius on (a,c,e,g,i) a
        metal-dielectric multilayered substrate
        ($\epsilon_{dielectric} = 3.1$, see the inset in (a) for
        $\epsilon_{silver}$) and (b,d,f,h,j) an effective homogeneous
        medium with the effective permittivity as in
        Fig. \ref{material_dispersion}: (a,b) scattering cross-section
        spectra, (c,d) the summary of the resonant wavelengths, the
        scattering cross-section normalised to the geometric
        cross-section, and electric $p_y$ and magnetic moments $m_z$,
        (e-j) the spatial distribution of the electric field
        amplitudes at resonant wavelengths (normalised to the incident
        field).}
      \label{md_vs_emt}
\end{center}
\end{figure}
\clearpage
\section{Conclusion and outlook}
In this paper, we performed comprehensive numerical studies of the
substrate influence on the optical properties of high-index dielectric
nanoparticles. Different types of substrates such as flint glass,
perfect electric conductor, gold, and hyperbolic metamaterial were
investigated. Retardation effects were shown to play significant role
in the particle-substrate interactions making the ``classical'' tools
such as image theory to be of limited applicability even in the case
of a PEC substrate. Moreover, substrates supporting nontrivial
electromagnetic excitations such as surface plasmon polaritons in case
of plasmonic metals and high-density of states extraordinary waves in
the case of hyperbolic metamaterials, give rise to complex resonant
responses and open a possibility for on-demand tailoring of optical
properties. The presence of substrates was shown to introduce
significant impact on particle’s magnetic resonances and resonant
scattering cross-sections. As for overall comparison of different
substrates, we can observe that dispersionless dielectric substrates
are the best for preserving natural properties of individual
particles, PEC and metal-dielectric substrates give a relatively
strong the MQ resonance and the best enhancement of the ED resonance,
and gold substrate is beneficial for the MD resonance. Variation of
substrate material provides an additional degree of freedom in
tailoring properties of emission of magnetic multipoles and designing
Fano-like resonances combining magnetic and electric excitations.

\nolinenumbers %%% do not use line numbers any more.
\section{Acknowledgements}
This work was supported, in part, by the government of the Russian
Federation (Grant 074-U01 and 11.G34.31.0020) and EPSRC (UK). AZ
acknowledges support from the Royal Society and the Wolfson
Foundation. Authors acknowledge discussions with Prof. Yuri~S.~Kivshar
and Dr. Andrey~Miroshnichenko.
\renewcommand{\refname}{References}

\end{document}